\documentclass[12pt]{article}
\usepackage{helvet}
\usepackage[round]{natbib}
\usepackage{amsmath}
\usepackage{amssymb}
\usepackage{amsthm}
\usepackage{dsfont}
\usepackage[margin=1in]{geometry}
\usepackage{hyperref}
\usepackage{multirow}
\usepackage{color}
\usepackage{tikz}
\usepackage{graphicx}
\usepackage{setspace}
\usepackage{algorithm2e}
\usepackage{float}
\usepackage{lscape}
\usepackage{rotating}
\usepackage{longtable}
\usepackage{subcaption}
\usepackage{etoc}
\usepackage{authblk}
\usepackage{ulem}

\newtheorem{assumption}{Assumption}
\hypersetup{
    colorlinks=true,
    linkcolor=black,
    filecolor=magenta,
    urlcolor=blue,
    citecolor=black,
}

\newcommand{\indep}{\mathrel{\text{\scalebox{1.07}{$\perp\mkern-10mu\perp$}}}}

\newcommand{\E}{\mathbb{E}}

\newcommand{\W}{\mathbf{W}}
\newcommand{\OCD}{\Omega_{\mathcal{C}, D}}

\newcommand{\PS}{PSM}
\newcommand{\GM}{GM}
\newcommand{\bX}{\mathbf{X}}
\newcommand{\bU}{\mathbf{U}}
\newcommand{\bx}{\mathbf{x}}
\newcommand{\bu}{\mathbf{u}}

\newcommand{\tauhat}{\hat{\tau}}

\begin{document}

\addtocontents{toc}{\protect\setcounter{tocdepth}{0}}

\title{Matching Bounds: How Choice of Matching Algorithm Impacts Treatment Effects Estimates and What to Do About It.}
\author[1]{Marco Morucci} 
\author[2]{Cynthia Rudin}
\affil[1]{Center for Data Science, New York University}
\affil[2]{Department of Computer Science, Duke University}
\author[]{}
\date{}
\maketitle
\thispagestyle{empty}
\begin{abstract}
Many major works in social science employ matching to make causal conclusions, but different matches on the same data may produce different treatment effect estimates, even when they achieve similar balance or minimize the same loss function. We discuss reasons and consequences of this problem. We present evidence of this problem by replicating ten papers that use matching and we find that different popular matching algorithms produce inconsistent results. We introduce Matching Bounds: a finite-sample, nonstochastic method that allows analysts to know whether a matched sample that produces different results with the same levels of balance and overall match quality could be obtained from their data. We apply Matching Bounds to a replication of two studies and show that in one case results are robust to this issue and in another they are not. 
\end{abstract}

\clearpage
\setcounter{page}{1}
\section{Introduction}

Quantification and reduction of uncertainty surrounding empirical results is a central pillar of political science, and of social science generally. Results accompanied by estimates of the statistical and non-statistical uncertainty that surrounds them are more reliable, credible, and lead to substantially stronger theoretical conclusions as well as better downstream policy applications. 

Despite the fact that the field specifically seeks to quantify (or eliminate) sources of uncertainty that could influence results, there is a popular type of technique for which a large amount of uncertainty goes unmeasured: \textit{matching}. Matching is widely used in applied political science due to its ability to produce nonparametric treatment effect estimates in an interpretable way, and many established empirical results have been obtained with matching methods. In addition, recent years have seen a proliferation of different matching methods~\citep[e.g.,][]{Rosenbaum1983,Diamond2013,Iacus2012,King2017}, all of which have the same desirable features that any good matching procedure should have, and,  importantly,  produce similar levels of balance on the same data: there is no clear way to choose one method over another. Choosing one method without taking others into account ignores the existence of potentially many good match assignments, potentially leading to substantial unquantified uncertainty in the resulting estimates. One might hope that the choice of matching method is innocuous -- no matter which matching method is used, as long as identification assumptions and balance constraints are satisfied the conclusion is always the same, but as we will show, this is not true.

In this paper, we show that many nominally equivalent matching methods produce different treatment effect estimates \textit{and support different conclusions on the same datasets}. We demonstrate this by replicating ten papers that conduct their analysis with matching, and by varying which matching method is employed. All of the papers we replicated have appeared in the American Journal of Political Science, and their results and conclusions are potentially influential in policy-making. We find that different popular matching methods lead to different conclusions. We argue that this issue is caused purely by the existence of many good match assignments among the data that produce similar balance (or otherwise measured quality) but differ in the estimates they lead to. The fact that many equally-good match assignments exist is due to inherent randomness in the data, implementation choices and other researcher degrees of freedom inherent in the use of matching; all of these are types of uncertainty that scientists typically would want quantified (but here, are not).  

The issue of different estimates from equivalently good matches has been largely ignored in the applied literature in political science: a survey of 100 papers that appeared in Political Science journals and that use matching methods was conducted by \cite{Miller2015}, who showed that only 9 of these papers reported results obtained with more than one matching method, and when they did, it was usually two. Virtually all of these studies make some reference to balance, or use the fact that the matching method they employ achieves good balance to justify its use alone \citep{Miller2015}. However, as we will show, even adding balance constraints does not eliminate uncertainty due to the choice of matching method.

To illuminate the uncertainty due to matching, we introduce Matching Bounds: a matching algorithm based on principles from robust optimization that outputs sets of good matches spanning the entire range of treatment effect estimates that could be produced by any reasonable analyst (where a reasonable analyst works within a specific set of constraints). Using Matching Bounds, analysts will be able to discover the largest and smallest treatment effect estimates that could be produced by changing their matching algorithm, which quantifies the uncertainty due to matching and thus leads to more robust inferences. Simply stated, Matching Bounds makes uncertainty stemming from choice of match assignment explicit and transparent. 

Matching Bounds works by solving two mixed-integer programs that respectively maximize or minimize the value of an average treatment effect estimator on the same data by changing only which units are matched with the constraints that the matches must produce the same level of balance, both global and covariate-specific, and measured in a variety of ways. In this way, Matching Bounds captures both statistical uncertainty and researcher-degrees-of-freedom.

Matching Bounds can be particularly useful when the robustness of empirical results is paramount: many studies in political economy and development economics have gone on to inform both the general theoretical frameworks of political science and the policy-making of major development agencies and state bureaucracies. For example, \cite{Clayton2015} uses matching to evaluate mandatory gender quotas in Lesotho, \cite{Zucco2013} uses matching to show that a conditional cash transfer intervention impacts electoral outcomes, and \cite{Buntaine2015} uses the same methods to argue that civil society organizations can hold international organizations such as the World Bank accountable to national authorities. All of these studies, whose data were used in our experiments, have great relevance for building and testing political science theories that go on to inform policy-making. Therefore, it is important that more robust matching methods are developed to strengthen the results presented by studies such as these. The method proposed in this paper is one such robust approach.

To further test and validate Matching Bounds, we will apply the methodology to a replication of two different studies that originally use matching. The first is a study of whether violence incites migration originally conducted by \cite{adhikari2013conflict}: here we show that results obtained with matching are largely robust to variation of match assignment at similar levels of balance. The second is a study of whether countries with more left-leaning governments tend to impose trade restrictions \citep{grieco2009preferences}: in this case we find that the results of matching analyses on these data are not robust to variation of match assignment. This means that there exist several same-balance matched subsets of the data that lead to largely different effect estimates.


The paper will proceed as follows: First we explain why matches of similar quality can lead to different results, and we provide empirical evidence that this occurs in practice: we replicate the analyses of ten prominent papers and show that their results may change if we switch from one popular matching method to another. Second, we introduce the Matching Bounds algorithm that computes the range of sample average treatment effects over reasonable match assignments. Finally, we apply Matching Bounds to the two studies of \cite{adhikari2013conflict} and \cite{grieco2009preferences} in Section \ref{sec:appl}.

\subsection{Matching Methods for Observational Inference: Definitions, Notations and Assumptions}\label{Sec:Notation}
We have a sample $D=\{(y_i, t_i, \mathbf{x}_i)\}_{i=1}^N$ of $N$ units, that we wish to study under a binary treatment, $t$. We measure an outcome variable of interest, $y_i$ for each unit in the sample, as well as a vector of $P$ covariates, $\mathbf{x}_i = (x_{i1}, \dots, x_{iP}) \in \mathbb{R}^{P}$, with the corresponding $N\times P$ matrix of covariates denoted by $\mathbf{X}$. We will most often refer to the two groups of units separately, denoting the treatment group with $\{(\mathbf{x}^t_i, y^t_i)\}_{i=1}^{N^t}$ and the control group with $\{(\mathbf{x}^c_j, y^c_j)\}_{j=1}^{N^c}$, where $N^t$ is the number of units that have received the treatment and $N^c$ is the number of units that are in the control group. 

In a common causal inference setting, we wish to estimate a treatment effect for each unit: $ \tau_i = Y_i(1) - Y_i(0)$, where $Y_i(1)$ and $Y_i(0)$ are unit $i$'s potential outcomes, namely, the values of $Y$ obtained under two different treatments. Unfortunately, we only ever observe one of the two potential outcomes for each unit and never the other.

Formally, we define match assignment as a matrix, $\W \in \mathbb{R}^{N^t \times N^c}$, such that every entry obeys $0 \leq w_{ij} \leq 1$.
The different types of matches that have been suggested in the literature can be obtained by constraining $\W$ in different ways.\footnote{For example: if we want one-to-one matching we require: $\sum_{i=1}^{N_t} w_{ij} = 1, \sum_{j=1}^{N_c} w_{ij} = 1, w_{ij} \in \{0,1\}$ for all $i,j$; if we instead want to allow multiple control units to be matched with the same treatment unit but no control unit can be matched more than once we require: $w_{ij} \in \{0,1\}$ and $\sum_{i}^{N_t}w_{ij} = 1$ for $j=1, \dots, N_c$. 
Another example: synthetic control matching \citep{Abadie2010} tries to average multiple control units to create a synthetic match for each treatment unit and can be obtained by requiring: $\sum_{j=1}^{N_c}w_{ij} = 1$ and $w_{ij} \in [0,1]$ for all $i$ and $j$.
} 
A matching method is a function: $A: \mathcal{D}\mapsto \mathcal{W}$, where $\mathcal{D}$ is the domain we draw the data from and $\mathcal{W}$ is the set of all possible matrices of matches. 
In this article we focus on integer matching with and without replacement: a single unit can be matched to one or more whole other units.
\footnote{While all matching estimators are weighting estimators~\citep{Imbens2015}, we choose to focus specificially on integer matching and not weighting in general. Similarly, we do not study matching as a tool for blocking before randomization, but only as a method for observational inference.} 

We make the common assumptions of observational inference. 
For all units, $i$, we assume:
\begin{assumption}{(Common Support)}\label{CS} $0 < \Pr(T_i =1|X_i) < 1$ \end{assumption}
\begin{assumption}{(Random Sample)}\label{SUTVA} $(Y_i(t), T_i, X_i) \overset{iid}{\sim} f(Y_i(t), T_i, X_i)$ \end{assumption}
\begin{assumption}{(Conditional Ignorability)}\label{ITA} $Y_i(1), Y_i(0) \indep T_i |X_i$.\end{assumption}
The first assumption requires all units to have a positive probability of being treated or untreated. The second assumption states that there is a well-defined population distribution for treatment, outcomes, and covariates, and that all units are drawn independently from it~\citep[see, e.g.,][]{Abadie2006}. Note that, formulated in this way, this assumption also indirectly requires the common SUTVA \citep{Rubin1976}, i.e.,  the absence of hidden treatments and the independence of units' outcomes from other units' treatment assignments. 
The last assumption states that all potential confounders of the causal relationship between $T$ and $Y$ are measured~\citep{Imbens2015,Pearl2009}. Throughout most of the paper, we will be concerned with the problem of estimating the Sample Average Treated Effect on the Treated (SATT) under the assumption above: 
\begin{align}
\tau &= \frac{1}{N^t}\sum_{i=1}^{N^t}\E[Y_i(1) - Y_i(0)|T_i=1, X_i=x_i].\label{Eq:SATT}
\end{align}
We choose to estimate this with the common difference-in-means matching estimator, defined for  match assignment $\W$ as \citep{Rosenbaum1983}:
\begin{align}
\hat{\tau} = \frac{1}{N^t}\sum_{i=1}^{N^t}y_i^t - \frac{1}{N^t}\sum_{j=1}^{N^c}y_j^cw_{ij},\label{Eq:ATTEst}
\end{align}
where recall that $w_{ij}=1$ when $i$ is matched to $j$.

\section{How and Why Changing Matching Methods Changes Treatment Effect Estimates}\label{sec:howandwhy}

Analysts are often interested in estimating an average causal effect with matched data that satisfies some pre-defined balance requirement. Consider the problem of having to match $M$ units, that is, find $\W$ such that aggregate imbalance between the treated and control sample is either below or equal to some level $B$ (which could also be the minimum achievable value in the data), as measured by a difference in means between treated and control groups, for each covariate, $p$. That is,
\begin{eqnarray}\nonumber
\textrm{find }\W\;&\textrm{ s.t. }&\;
\left|\frac{1}{M}\sum_{i=1}^{N^c}(\W^T\mathbf{X}^t_p)_{i} - \frac{1}{M}\sum_{i=1}^{N^t}(\W\mathbf{X}^c_p)_{i}\right| \leq B \;\;\forall\; p, \;\textrm{ (balance constraint) }\\
&\textrm{ and }&\sum_{k}W_{k,l}=M \; \forall\; l, \; \sum_{l}W_{k,l}=M\; \forall\; k\;\; \textrm{ (at most $M$ matches per unit). }
\label{Eq:SystemBalance}
\end{eqnarray}
This is a system of inequalities in $N^t\times N^c$ unknown integer variables, 
and the analyst may choose to add additional constraints. (There is a constraint on the domain of $w_{ij}$ as well.) This problem clearly may have multiple solutions \citep{Papadimitriou1984,Valiant1986}, and there is no general way to devise balance constraints that ensure uniqueness of the solution of the system for an arbitrary dataset. This would be true even if we were to replace the inequality in \eqref{Eq:SystemBalance} with strict equality. 
If we allow only one-to-one matches, or even one-to-many in certain cases, or even replace the inequality with an equality, different solutions will match the same treatment unit to different control units.\footnote{It is important that the dependent variable never actually be used to make matches \citep{Rubin1976}, as this would equate with selecting cases on the dependent variable, which would introduce confounding in the design.}

Since we ignore the value of $Y$ completely when making matches, we also ignore the possibility that two units that have similar values of $X$ could have different values of $Y$; this is uncertainty that we do not want to inadvertently exclude. 

\subsection{Why This Happens}\label{sec:why}
\paragraph{Statistical reasons.} There are precise statistical reasons for why there might exist multiple sets of matches that lead to different treatment effect estimates. To study the most general setting, suppose Assumptions 1 and 2 hold. We have a real-valued random outcome $Y(t) = f_t(\bx, \bu, \epsilon)$, where $\bx$ represents the observed covariates, $\bu$ is a vector of \textit{unobserved} covariates, and $\epsilon$ is a real-valued scalar term denoting unrelated noise, such that $\E[\epsilon] = 0$. Consider two match assignments $\W^a$ and $\W^b$: we wish to estimate the SATT with the estimator in \eqref{Eq:ATTEst}. Let $\tauhat_a$ be the estimate of the SATT obtained with $\W^a$ and $\tauhat^b$ the estimate obtained with $\W^b$. We can use a simple Taylor expansion to decompose the difference between $\tauhat^a$ and $\tauhat^b$ as:
\begin{align}
\tauhat^a - \tauhat^b &= \underbrace{\boldsymbol{\beta}_1^T(\bar{\bx}^a - \bar{\bx}^b)}_{\text{Observables}} + \underbrace{\boldsymbol{\beta}_2^T(\bar{\bu}^a - \bar{\bu}^b)}_\text{Unobservables} + \underbrace{\beta_3(\bar{\epsilon}^a - \bar{\epsilon}^b)}_{\substack{\text{Baseline}\\\text{variance}}}\label{Eq:LinApprox}\\
&+ \text{difference of higher order terms},\nonumber
\end{align}
where $\bar{\bx}^a$ is the average value of observed covariates in matched sample $a$, $\bar{\bu}^a$ is the average value of unobserved covariates in matched sample $a$, $\bar{\epsilon}^a$ is the average value of statistical noise in matched sample $a$, and $\bar{\bx}^b, \bar{\bu}^b, \bar{\epsilon}^b$ are analogously defined for matched sample $b$. See the supplement for a full derivation and precise definition of all quantities. 
Clearly, the decomposition in \eqref{Eq:LinApprox} shows that there could be different reasons why $\tauhat^a - \tauhat^b$ is nonzero.

First, there is the difference in the averages of observed covariates: $(\bar{\bx}^a - \bar{\bx}^b)$. This difference is 0 whenever both $\W^a$ and $\W^b$ lead to perfect mean balance, i.e., for some metric $\|\cdot\|$ on the space of $\bx$, we have: $\|\bar{\bx}^a - \bar{\bx}^t\| = \|\bar{\bx}^b - \bar{\bx}^t\| = 0$, where $\bar{\bx}^t$ is the average of the observed covariates among treated units. Note that when $\|\bar{\bx}^a - \bar{\bx}^t\| = \|\bar{\bx}^b - \bar{\bx}^t\| \neq 0$, it is not necessarily true that $\bar{\bx}^a = \bar{\bx}^b$, and, therefore, the first term in \eqref{Eq:LinApprox} will generally be non-zero. This happens whenever the same level of mean balance, according to some pre-defined metric, is achievable by balancing different sets of covariates. 

The second term in \eqref{Eq:LinApprox} represents a weighted difference in average unobservables between the two matched control sets. This quantity is 0 under Assumption 3, or under the same conditions on mean balance as we had for observables. Importantly, balance can never be checked in this case because $\bu^a$ and $\bu^b$ are not observed. This suggests that the existence of different sets of matches of equal balance on observables, but that produce different estimates, could be an indicator of violation of Assumption 3. 

Lastly, the final term in \eqref{Eq:LinApprox} is the difference between residual terms of outcome variables. This quantity is guaranteed to be 0 only if there is no randomness at all in the outcomes, conditional on treatment and observed covariates, and under Assumption 3. This is rarely the case, as outcomes are often sampled from a super-population of interest. A simple upper bound on the expected absolute value of this term is given by $\sqrt{\textit{Var}(\bar{Y}^a|\bX, \bU) + \textit{Var}(\bar{Y}^b|\bX, \bU)}$:\footnote{See the supplement for a derivation.} this indicates that when outcomes have large conditional variance, then there is a possibility that units with similar values of both observed and unobserved covariates will still have different outcomes. Because of this, units with the same covariate values could have different outcomes under Assumption 3, adding another explanation of why there might exist multiple equally-balanced matched sets that lead to different treatment estimates. 

\paragraph{Researcher degrees of freedom.} There is also one fundamental non-statistical factor that influences the issue presented here: researcher degrees of freedom inherent in the choice of matching procedure, its hyperparameter values, and even the choice of balance criteria \citep{gelman2013, coker2018}. There is often little to no guidance on how to choose either a matching procedure or hyperparameter values. For example, the value of $B$ in \eqref{Eq:SystemBalance} must be chosen prior to analysis, or must be adjudicated as sufficient if output by some optimization algorithm. Moreover, different matching methods constrain the problem in \eqref{Eq:SystemBalance} in different ways, and analysts could choose to add or modify additional constraints on the problem. 
Combined with the statistical issues just presented, this amount of choice that researchers have in choosing and using a matching method only adds to the uncertainty caused by the existence of many good quality match assignments.

\subsection{Consequences}
\paragraph{Choosing the optimal-balance matches does not solve the problem.} 
As we have now shown, matched sets with similar quality but different average outcomes might still exist in the data due to both the statistical reasons highlighted above, and to researcher choice of optimization criteria, algorithms, and hyper-parameters. In particular, Assumptions 1-3 being satisfied does not guarantee the existence of a single optimal set of matches. 

Even if there were only one optimal-balance matched set on the specific data sample observed, the argument that this and only this set should be chosen is fallacious as it is predicated on the assumption that results that have exactly this balance should be trusted and nothing else, i.e., no match assignment that has a slightly different balance would be trusted, and no other slightly different definition of balance should be used. 
In this sense, choosing optimal balance is a strong assumption: a more robust approach would be to consider matched sets that are within a  threshold of the ``best'' quality so that results are not sensitive to small changes in data or researcher choices. 

\paragraph{Different matching methods produce different treatment effect estimates.} 
Another key consequence of the problem introduced before, and the main way in which it affects reliability of results, is that different algorithms will choose between several different equal-quality solutions differently from one another, and, in most cases, arbitrarily. This will cause different matching algorithms applied to the same dataset to produce different estimates. Obviously hyperparameter choice and the researcher-degrees-of-freedom that this involves play a fundamental role in why different matching methods would produce different estimates.
These often arbitrary choices have a great impact on how and why different procedures lead to different results. Nonetheless, the arguments made in this section should make it clear that even if we were to ignore the issue of researcher choice, many different, good quality, matched groups could still exist, and algorithms could be choosing among them in arbitrary ways. The observation that different matching methods lead to different results is surprisingly scarce in the existing literature: some existing work \citep{Miller2015,Noor2017} notices the problem and propose tentative solutions, but do not provide empirical or simulated evaluation of its magnitude.

\subsection{Empirical Evidence From Replicating the Analyses of 10 Papers}
A consequence of the fact that many similar-quality matched groups exist is that different matching methods produce different estimates on the same data. We present empirical evidence of this fact in this section. We show results from a replication study of ten papers that appeared originally in the American Journal of Political Science after 2010. These papers study a range of different topics but all have in common that they make use of matching methods either for their primary hypothesis test or for a robustness check. The papers replicated here were chosen to represent a broad array of different areas of inquiry within political science: three are about topics in political economy and comparative politics, four are about American politics and the US party systems and institutions, and three are about international relations and  civil conflict. Within each subfield, papers were selected based on two criteria: data size and availability and the extent to which matching was used within the paper. 

Crucially, all but one of the papers report results from only one single matching method, thus ignoring the possibility that results might have been different had they chosen to do their matches differently. Here we tried to replicate as many of the hypothesis tests on the paper's primary treatment effect, reported by the authors. The existence of more than one treatment effect is tested in each paper, and different outcome models are used to estimate effects after matching. We tried to replicate as many of the tests reported in each paper as possible, changing estimands or outcome models as the authors do.

Expanding the analyses of the original papers, we used 6 popular matching methods and compare their results. We compared matches made using Nearest Neighbor, Coarsened \citep{Iacus2012}, Optimal \citep{Rosenbaum1989}, Genetic \citep{Diamond2013} and Mixed Integer Matching \citep{Zubizarreta2012}. We chose these algorithms because of their popularity in applied research,\footnote{Of around 80 of the Political Science papers that make use of matching methods surveyed by~\cite{Miller2015}, 19 make use of CEM, 28 of Genetic Matching, and 38 Nearest Neighbor matching. Optimal and MIP matching are less used in political science but are largely employed in biostatistics and epidemiology.} and because there are large differences between how they choose which matches to use among equivalently good ones or how they assess balance. While different, all of these methods have been shown capable to achieve balance within their samples. Additional information on how this replication was conducted is available in Section \ref{Appendix:Replication} of the Supplement.

After creating a matched dataset with each of the six methods chosen, we applied the exact estimation method used by the papers' original authors on their matched data. This was almost always a linear regression model, the coefficients of which are interpreted as treatment effect estimates. Pre-estimation matching methods all target the SATT. 

\begin{sidewaysfigure}[!htbp]
  \begin{minipage}[t]{0.45\textwidth}
    \caption{Consistency of treatment effects under different matching methods.}
    \includegraphics[width=1.2\linewidth]{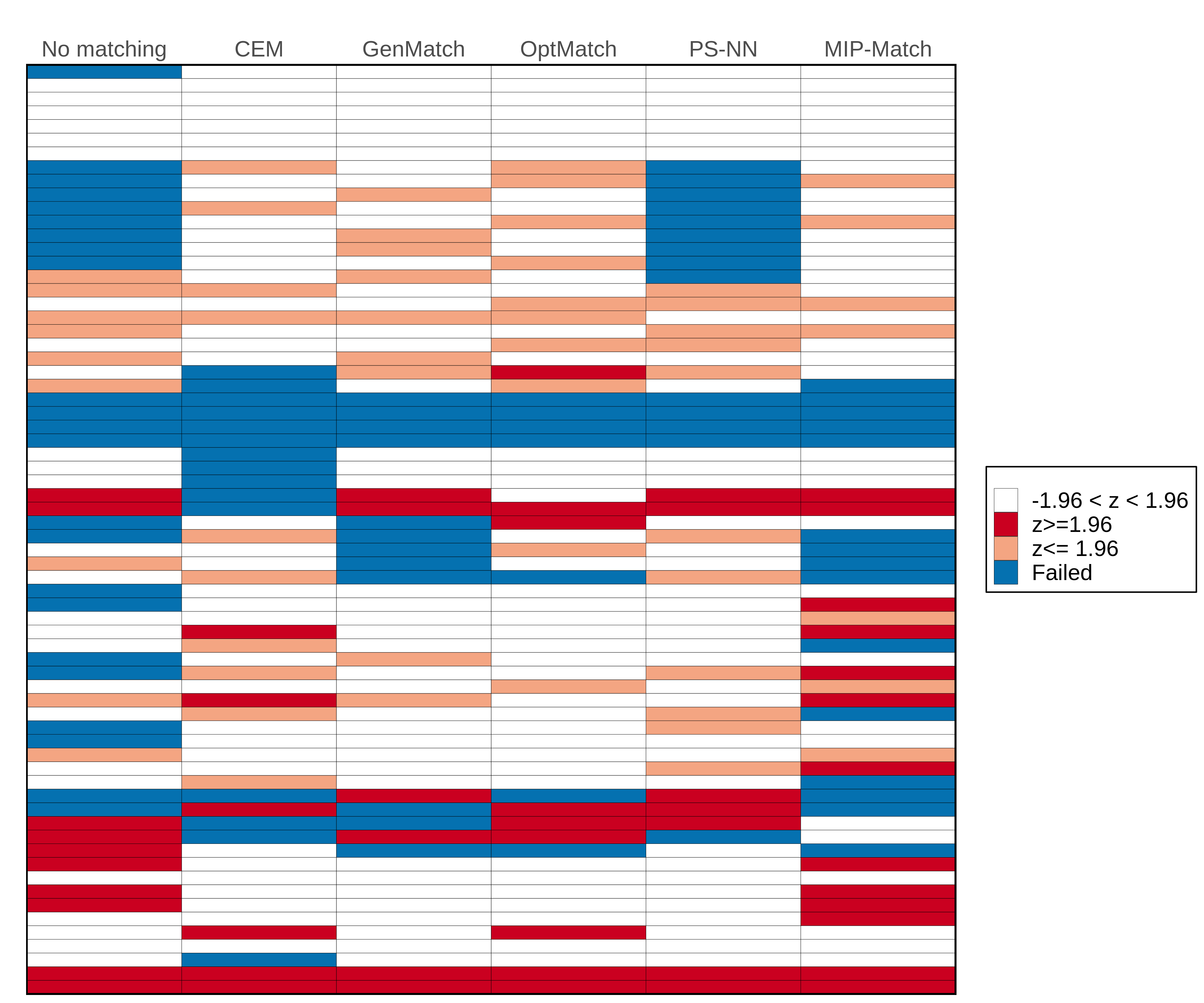}
     \footnotesize{Note: Each row is a hypothesis test conducted in one of the papers (multiple tests are conducted by each paper) and each column a different matching algorithm after which the test was performed. Cells are colored according to the result of the test. When tests are coded as failed, it is because either the matching procedure or the subsequent model failed to run under the specification provided by the original author. A numeric version of this figure is available in Table~\ref{A:Res1} of the Supplement.}
\label{Res1}
  \end{minipage}
  \hfill
  \begin{minipage}[t]{0.45\textwidth}
    \caption{Consistency of covariate balance under different matching methods.}
    \includegraphics[width=1.2\linewidth]{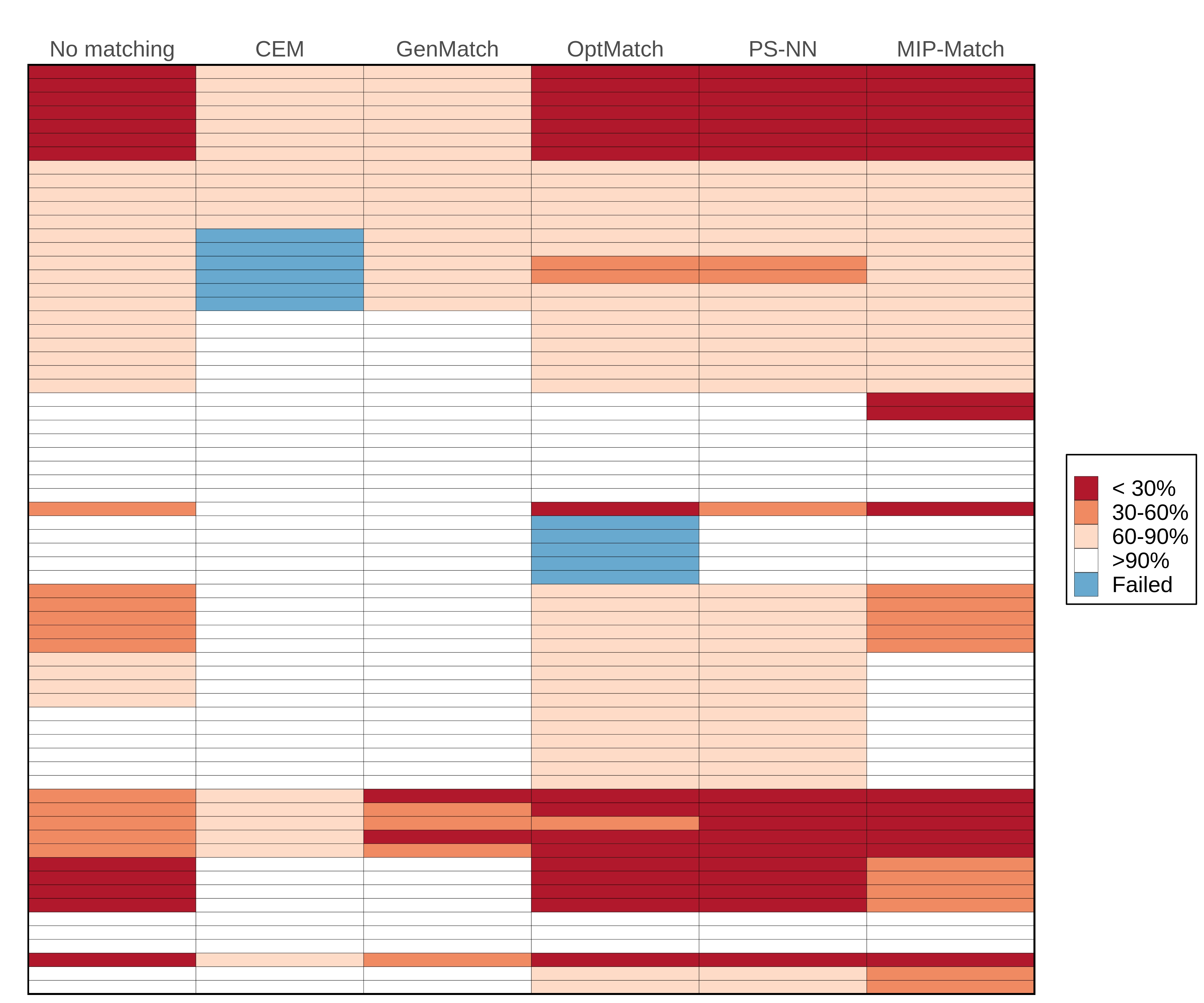}
    \footnotesize{Note: Each row is a hypothesis test and each column a matching algorithm. Cells are colored according to the percentage of covariates balanced by the matching method. This is calculated as the number of covariates whose standardized absolute difference in means between treated and control is below 0.5 after matching. When tests are coded as failed, it is because the matching procedure failed to run with the original authors' set of covariates. A numeric version of this figure is available in Table~\ref{A:Bal1} of the Supplement.}
\label{Bal1}
  \end{minipage}
\end{sidewaysfigure}

Results are reported in Figure~\ref{Res1}. Each row in the figure corresponds to a hypothesis test (multiple hypothesis tests are performed in each paper) and each column to a matching method, and cells are color-coded based on the result: results that are positively significant at the 5\% level are red, non-significant results are white, and negatively significant results at the 5\% level are orange.  In some cases, post-matching estimators fail to run on the matched data produced by one of the methods:\footnote{In most of these cases, regression models based on a numerical optimization method failed to converge to a solution on the matched data.} we denote this with a blue cell. We chose not to alter the post-matching method in these cases to remain as close as possible to the authors' original analysis. \textit{If choice of matching algorithm did not influence results, we would see rows colored uniformly, but this is not the case in the figure: different matching methods produce different results.} 

Figure~\ref{Bal1} reports balance statistics for the same set of matching methods in a similar way: cells are color coded according to what percentage of covariates was balanced after matching with each method. \textit{Importantly, even though balance consistency across matching methods does vary, its variation is not as large as that of the hypothesis test results.} 

\textit{For about 90\% of the replicated hypothesis tests, at least one matching method produced a statistically significant result and another produced a non-significant result.} This is of substantive importance because, while statistical significance is not in itself an indicator of the substantive significance of the results, it is often a heuristic employed to evaluate observational inference results. Moreover it is undeniable that publication bias favors statistically significant results, which will then have a wider readership and potential impact. For about 62\% of the hypothesis tests conducted, there is at least one matching method that produces a positive treatment effect and one that produces a negative one. Given that in most cases the methods achieve similar balance, this makes it hard to discern even the direction of the treatment effect.

Our replication exercise shows that there is more uncertainty in empirical results produced with matching methods than commonly thought. This additional uncertainty is caused by choice of matching method, as evidenced by the data presented here. 

\section{Quantifying Matching Uncertainty with Matching Bounds.}
In this section, we introduce Matching Bounds: a method for making variation in treatment effect estimates due to matching explicit to the analyst, which mitigates the issues discussed earlier. We propose a set of algorithms aimed at computing the smallest and largest treatment effect (TE) estimates that could be obtained on a set of data by simply changing how the matches are made subject to match quality constraints, such as balance or calipers. This approach is guided by principles of robust optimization \citep[see, e.g.,][]{Beyer2007}; if traditional matching methods solve an optimization problem to find the matched set of units that maximizes a certain criterion, then fluctuations both in the data and the parameters of the matching procedure can make the optimal solution unstable, as shown previously. Robust optimization as a field aims to deal with such instability problems in general optimization problems. We will apply robust optimization methodologies to the optimization problem solved by matching. In our case, we consider that an analyst could choose any reasonable match assignment arbitrarily, and we examine the range over reasonable match assignments. This range is called the \textit{uncertainty set} in robust optimization. We quantify uncertainty as usual in robust optimization by calculating the range of results over the uncertainty set.



Let us define Matching Bounds formally. Let $\mathcal{C}$ be a set of constraints on the quality of the matches. These could be, for example, calipers, balance constraints or aggregate distance constraints. Now let $\OCD$  be the set of matches that satisfies all the constraints in $\mathcal{C}$. The matching upper and lower bound for a generic estimator, $\hat{\theta}$, are:
\begin{align}
\W^+ \in \arg\max_{\W \in \OCD} \hat{\theta}(\W, D),\quad
\W^- \in \arg\min_{\W \in \OCD} \hat{\theta}(\W, D).
\end{align}
In words, these bounds are the largest and smallest estimates that could be obtained for $\hat{\theta}(\W, D)$ by changing the matches within the feasible region. Using these matches, analysts can compute $\hat{\theta}(\W^+, D)$ and $\hat{\theta}(\W^-, D)$:  the range of treatment effects that could be obtained on the same sample, and with the same level of balance, by changing how matches are made. Knowing this information is important for both scientific and policy purposes: In the first case, Matching Bounds allows analysts to test scientific theories independently of the matched sample used. In the second case, our proposed method can be used to bound average effects of policies also independently of matched sample. Practically, Matching Bounds highlights the potential variation in estimated effects at similar levels of match quality: if two matching algorithms produce similar levels of match quality, TE estimates they produce will lie between the bounds. 

There are several other reasons why this approach is advisable over alternatives. 1) \textit{Computing all equally good matches and then averaging treatment effects over them may seem like a better solution than bounding the range of these effects, however it is not.} Averaging estimates from different matches presumes that the probability of choosing a specific assignment be known and uniform. Since this probability depends both on choice of a matching method and on the matching method's choice of assignment, this quantity is unlikely to be even close to a uniform distribution over reasonable match assignments. The match assignment acquires bias from the analyst's chosen matching algorithm, the order that the data are provided to it, and in other ways. Thus, we should never assume that a match assignment would be chosen uniformly. 2) \textit{Even if the choice of match assignment were chosen uniformly at random by the analyst (which it is not), the randomness from this choice should be included in quantification of uncertainty.}
Most estimators used after matching do not account for this variance, leading to biased variance estimation. Moreover, enumerating all possible good matches is a computationally expensive problem, as it involves solving a large sequence of increasingly complex integer programs.\footnote{We might solve the problem, then add a constraint that the previous solution should not be chosen, and re-solve. We might repeat this many times to enumerate all solutions. This is a sequence of increasingly complex integer programs.}

While the bounds above are defined as extrema of a function, they do not represent outlying quantities: \textit{all} feasible match assignments in the optimization problem are reasonable match assignments that an analyst might choose, defined by the constraints in $\mathcal{C}$. 
Matching Bounds are searched only among those matches that satisfy these requirements. An analyst wanting to estimate Matching Bounds for a dataset should choose these constraints based on the minimal requirement for quality that she would want for the matches with any methodology. Matches made at the bounds should be indistinguishable in terms of quality from any other good set of matches. Even if there were no popular matching algorithms that produced the same treatment effect as the Matching Bounds, it would still be important to know that at the same level of match quality, there existed matches that produced such different TE estimates. 

\subsection{Computing Matching Bounds for the SATT}
We now outline computational strategies to obtain Matching Bounds for two different estimands of interest under Assumptions 1, 2, and 3. 
We would like to compute Matching Bounds for the SATT as defined in Eq. \eqref{Eq:SATT}, using the estimator in \eqref{Eq:ATTEst} with $M$-nearest neighbor matching. If we constrain the number of matches to be exactly the number of treated units, then we can compute Matching Bounds on the SATT by solving the following integer program, letting $\bar{y}^t = \frac{1}{N^t}\sum_{i=1}^{N^t}y_i^t$:\\
\textbf{Formulation 1}
\begin{subequations}
\begin{align}
\mbox{min/max } &\hat{\tau} = \bar{y}^t - \frac{1}{MN^t}\sum_{i=1}^{N^t} \sum_{j=1}^{N^c}y_j^cw_{ij},
\end{align}
\begin{align}
\mbox{subject to: }&w_{ij} \in \{0,1\},&&\quad i = 1,\dots,N^t,\; j=1,\dots,N^c \label{c1.1} \\
&\sum_{i=1}^{N^t}w_{ij} \leq K^c,&&\quad j=1,\dots, N^c \label{c1.2}\\
&\sum_{j=1}^{N^c}w_{ij} = M,&&\quad i=1,\dots, N^t\label{c1.3}\\
&\W \in \OCD. &&\label{c1.4}
\end{align}
\end{subequations}
This integer program targets the treatment effect as its objective, and constrains matches to be integer~\eqref{c1.1}, and each control unit to be matched at most $K^c$ times~\eqref{c1.2}. One-to-$M$ matching is enforced by Constraint~\eqref{c1.3}. Finally, additional constraints in \eqref{c1.4} are specified within set $\mathcal{C}$, and $\OCD$ is the set of all matches that satisfy the constraints in $\mathcal{C}$ on data $D$. Examples of constraints such as balance or aggregate distance are given in Section \ref{Sec:Constraints} of the supplement. Note that $K^c$ needs to be chosen to be sufficiently large as to allow for $M\times N^t$ matches. If the pool of control units is smaller than the pool of treated units, replacement must be possible in order for the SATT to be estimable in this manner. In this case, each treatment unit is matched to one control unit, but the same control unit can be used up to $K^c$ times.

\subsection{Computing Matching Bounds for Same-Balance-Subsample ATTs}\label{Computing}
Another quantity of interest is the range of treatment effects in subsamples of the data that are equally balanced or, more generally, satisfy the same constraints. That is, by allowing the exclusion or inclusion of certain treated units among the matched samples, we can find subsamples that are equally balanced, at the expense of not fully representing the whole treated sample. 
In some cases, treated units are excluded because of theoretical interest: for example, analysts might not want to include certain units that appear to be outliers with respect to the matching covariates. In some other cases, excluding some treated units from the sample is done out of necessity, as no balanced subset of controls can match to all treated units. Related to a point of \citet{Iacus2012, King2017}, we note that sub-sample Matching Bounds can be useful to characterize variation in matching estimates due to selection of such a sub-sample. In general, knowledge of the range of TE estimates between equally-good subsamples of the data should inform both empirical evaluation of theoretical premises and policy decisions that are based on empirical evaluation. 

For a chosen set of matches, $\W$, where treated units are not matched multiple times and are either included or excluded in the sub-sample, i.e., $\sum_{j=1}^{N^c}w_{ij} \leq 1$, we adopt the same terminology as \citet{King2017} and define the effect of interest as the Feasible Average Treatment Effect (FSATT): 
\begin{align}
\tau^s &= \frac{\sum_{i=1}^{N^t}\sum_{j=1}^{N^c}w_{ij}\E[Y_i(1) - Y_i(0)|X_i=x_i]}{\sum_{i=1}^{N^t}\sum_{j=1}^{N^c}w_{ij}}. \label{Eq:FSATTest}
\end{align}
Again, the estimator of choice is a difference-in-means on the data matched by $\W$:
\begin{align}
\hat{\tau}^s = \sum_{i=1}^{N^t}\sum_{j=1}^{N^c}\frac{(y_i^t - y_j^c)w_{ij}}{\sum_{i=1}^{N^t}\sum_{j=1}^{N^c}w_{ij}}.
\end{align}
We differentiate between computing bounds when the desired number of matches is pre-specified by the user and when only lower and upper bounds are placed on the total number of matches. Given a desired number of matches, $M$, lower and upper Matching Bounds can be obtained for the FSATT by solving the following integer program:\\
\textbf{Formulation 2:}
\begin{subequations}
\begin{align}
\max/\min\; \frac{1}{M}\sum_{i=1}^{N^t} \sum_{j=1}^{N^c}y_i^tw_{ij} - \frac{1}{M}\sum_{i=1}^{N^t} \sum_{j=1}^{N^c}y_j^cw_{ij},
\end{align}
\begin{align}
\mbox{subject to: }&w_{ij} \in \{0,1\},&&\quad i = 1,\dots,N^t,\; j=1,\dots,N^c \label{c3.1} \\
&\sum_{i=1}^{N^t}w_{ij} \leq K^c,&&\quad j=1,\dots, N^c \label{c3.2}\\
&\sum_{j=1}^{N^c}w_{ij} \leq 1,&&\quad i=1,\dots, N^t \label{c3.3}\\
&\sum_{i=1}^{N^t}\sum_{j=1}^{N^c} w_{ij} = M,&&\quad i = 1,\dots,N^t,\; j=1,\dots,N^c \label{c3.4}\\
&\W \in \OCD. &&\label{c1.5}
\end{align}
\end{subequations}
The formulation above is analogous to Formulation 1, except with two additional constraints: \eqref{c3.3} forces treated units to only be used once, and \eqref{c3.4} defines the overall number of matches, given by $M$. 

The formulation above allows analysts to compute Matching Bounds for the SATT under one-to-one matching. This latter requirement can be too strict: indeed many popular matching algorithms optimally choose the number of controls matched to each treated unit~\citep{Iacus2012, Diamond2013, King2017} to maximize balance. For this reason, we also introduce a formulation with an unspecified number of matches that requires only a lower and upper bound for the total numbers of times each unit can be used as a match. 
The objective of this formulation is fractional in the decision variables, making the optimization problem much harder to solve. In order to keep this optimization problem linear in the decision variables, we apply a Charnes-Cooper transformation \citep{Charnes1962} to the fractional objective to linearize it. To do this, we introduce the following auxiliary variables:
\begin{align}
u_{ij} = \frac{w_{ij}}{\sum_{i=1}^{N^t}\sum_{j=1}^{N^c}w_{ij}} \quad \mbox{and} \quad z = \frac{1}{\sum_{i=1}^{N^t}\sum_{j=1}^{N^c}w_{ij}},
\end{align}
The variable $z$ is the inverse of the number of control units that treated unit $i$ is matched to, and this number is now allowed to vary, but will be constrained to be nonzero. (This is equivalent to constraining at least one $w_{ij}$ to be nonzero in the sum in the denominator.) The variable $u_{ij}$ is a fractional version of the binary indicator that determines whether unit $i$ is matched to unit $j$. With these new decision variables we can formulate the problem of variable number matching as:\\

\textbf{Formulation 3}
\begin{subequations}
\begin{align}
\max/\min\;  &\sum_{i=1}^{N^t}\sum_{j=1}^{N^c}y_i^tu_{ij} - \sum_{i=1}^{N^t}\sum_{j=1}^{N^c}y_j^cu_{ij},
\end{align}
\begin{align}
\mbox{subject to: }& w_{ij} \in \{0,1\}, z, u_{ij} \in [0, 1]&&\quad i = 1,\dots,N^t,\; j=1,\dots,N^c \label{c5.0}\\
& u_{ij} \leq w_{ij}, &&\quad i = 1,\dots, N^t,\;j=1,\dots,N^c\label{c5.2}\\
& u_{ij} \geq z - (1-w_{ij}), &&\quad i = 1,\dots, N^t,\;j=1,\dots,N^c\label{c5.3}\\
& \sum_{i=1}^{N^t}\sum_{j=1}^{N^c}u_{ij} = 1, \label{c5.5}\\
& \sum_{j=1}^{N^c}u_{ij} \leq z, &&\quad i = 1,\dots,N^t\label{c5.6}\\
& \sum_{i=1}^{N^t}u_{ij} \leq K^c\cdot z, &&\quad j = 1,\dots,N^c\label{c5.7}\\
& \W \in \OCD.\label{c5.9}&&
\end{align}
\end{subequations}
The objective is the usual difference in sums between treated and control units included in the match. Constraint \eqref{c5.0} defines the domain of all the decision variables. Note that the auxiliary variables just introduced can have only non-negative values. Constraints \eqref{c5.2}, \eqref{c5.3}, \eqref{c5.5} define the role of the new decision variables, $u_{ij}$: these variables represent whether or not unit $i$ is matched to $j$, but they are not explicitly constrained to be binary variables, they are instead forced to be either zero (by Constraint \eqref{c5.3} when $w_{ij}=0$, or exactly $z_i$ (Constraints \eqref{c5.5} and \eqref{c5.2}) when $w_{ij} =1$. Constraint \eqref{c5.6} forces each treatment unit to be used only once, and Constraint \eqref{c5.7} forces each control unit to be used at most $K^c$ times. Additional match quality constraints are defined by \eqref{c5.9}. Note that when $K^c = N^t$ we have matching with full replacement, while $K^c = 1$ defines one-to-one matching.  In Section \ref{Sec:Constraints} of the supplement, we provide versions of constraints on $w_{ij}$ that are transformed to work with $u_{ij}$ as decision variables instead. We do so for some popular balance and match quality measures. 

The proposed optimization methods can be implemented with any of the popular Mixed-Integer-Programming tools such as GLPK, CPLEX or Gurobi. In certain cases, linear relaxations can also be considered if the analyst is willing to forgo unitary matches for computation time. More information on integer programming methods and tools is available in~\cite{Boyd2004} and \cite{Winston2003}.

\subsection{Practical Guidelines for Using Matching Bounds}\label{MBandMatching}

In this section, we give practical guidelines for using Matching Bounds as part of common data analysis pipelines in political science. 

\paragraph{Using matching bounds in conjunction with another matching method.} Matching Bounds can be used as a robustness check for matches produced by another matching algorithm. Given a desired matching algorithm, $A$ and a dataset, $D$, we can run that algorithm on the data to obtain a match assignment $\W^A$. Once that match assignment is obtained, desired match quality statistics can be computed on the matched data.\footnote{For example, aggregate distance between matches can be computed with $L^A = \sum_{i=1}^{N^t}\sum_{j=1}^{N^c}w_{ij}^Ad_{ij}$, while moment-balance for the $k^{th}$ moment of the $p^{th}$ covariate is computable after running $A$ with: $\sigma_{pk}^A = \bigl|\sum_{i=1}^{N^t}\sum_{j=1}^{N^c}w_{ij}^A[(x^t_{ip})^k - (x^c_{jp})^k]\bigr|$.} 
We can then use Matching Bounds to consider all match assignment of equal or better quality as those produced by $A$. For example, to calculate match quality for $\W^A$, we can plug $\W^A$ into any of the constraint equations in Section~\ref{Sec:Constraints} that we would use to define match quality. This gives us the right hand side for those constraints to use in Matching Bounds. 

We recommend allowing for some tolerance (maybe 5-10\%) between the quality level achieved by $A$ and the quality at which Matching Bounds are computed. Restricting the bounds to be computed at the exact level of match quality output by $A$ might be misleading: there could be other matching procedures that output matches of slightly lower but still admissible quality. By loosening the requirements on how close the bounds have to be to the quality of the matches made by $A$, we can ensure that an appropriate space of potential admissible matches is considered to produce bounds. 

\paragraph{Using matching bounds to search over hyperparameter values.}  Matching Bounds can be directly used to quantify uncertainty due to researcher degrees of freedom by outputting treatment effect bounds over ranges of potential hyperparameters of a matching procedure \citep[i.e., these bounds are hacking intervals over the choice of matching algorithm, see][]{coker2018}. Such hyperparameters might include calipers, weight put on covariates to compute balance, or other metrics of distance between units. To use Matching Bounds in this way, we can constrain a chosen hyperparameter (or set of hyperparameters) to be within a desired search range in one of the MIP formulations introduced in this paper, and use differences between bounds obtained with different hyperparameter values to assess sensitivity of matching to variations in the chosen hyperparameter. It may be desirable, in this case, to run Matching Bounds several times, with different ranges for the hyperparameter being tuned, to get a better picture of discontinuity points and general impact of the hyperparameter. 

\paragraph{Do not rely on Matching Bounds as sensitivity analysis for violations of conditional ignorability.} We remark that all the methods presented in the paper are reliant on Assumption 3 (conditional ignorability), and we caution the reader against relying on Matching Bounds for measuring unobserved confounding. As Eq$.$ \eqref{Eq:LinApprox} shows, differences in estimates at similar levels of balance could be due to baseline conditional randomness, or different ways in which balance is achieved by differentially weighting covariates, as well as signal from unobservables. The only scenario we can envision in which Matching Bounds could provide information about the effect of unobserved covariates is when the effect is small: observed covariates are highly predictive of the outcome, leaving little outcome variance unexplained. Even in this setting, analysts must be cautious that their models are not overfitting the data, and that the treatment effect is substantial enough with respect to their prediction error, before concluding that a small difference in MBs implies absence of unobserved confounding. \footnote{See \cite{cinelli2020} for sensitivity analysis tools that rely on this setting, in conjunction with linear outcome models.} 

\paragraph{We should report Matching Bounds results (along with our conventional results).} Lastly, there might be concerns on how to report and interpret Matching Bounds in conjunction with other common statistics. Matching Bounds can be reported alongside other common statistics of the desired treatment effect estimate. Overall, Matching Bounds should be used as a robustness check on existing matching results, in conjunction with theoretically driven justification of the choices of match quality metrics and values. 

\subsection{Matching Bounds are Not Too Conservative: Simulated Evidence}

We present simulation results that investigate whether MBs are capable of detecting a treatment effect when there is one in the true DGP, and whether they can fail to detect one consistently when there is not. Our results below show that MBs are substantially more robust than other matching methods when the TE is 0, but are not too conservative, and can detect a TE when there is one. 

For $N=1000$ observations divided in 100 strata, we first create a fixed observed covariate: $S_i \sim \textit{Discrete}_{100}(0.01)$ (i.e., a uniform distribution),  which determines the stratum that an observation falls in. Each stratum, $j$, is associated with a coefficient $\beta_{j}$ representing the strength of the covariate weight in the outcome and treatment models. We also fix a propensity score with: $e_i = \frac{1}{1 - e^{-\beta_{S_i}}}$. At each iteration of the simulation, we draw: $T_i \sim \textit{Bernoulli}(e_i)$, and generate outcomes $Y_i = T_i\times \tau + \beta_{S_i} + \epsilon_{i}$, where $\epsilon_{i} \sim \textit{Normal}(0, \sigma^2)$. Throughout our simulations, we vary two parameters: first, the value of the ATT, $\tau$, which we set to the values $0, 1, 3, 5, 8, 10$, second, the amount of conditional variance in outcomes, i.e., $\sqrt{\textit{Var}[Y(t)|S]}$, which we vary between 1 and 10. For each setting of the parameters, we perform 100 simulations. At each simulation, we perform a Z-test with the estimate of the ATT produced by each method tested. The standard deviation for these tests is computed by using the sample standard deviation of the estimates of $\tau$ output for each method over the hundred simulations. This ensures that standard errors are correct with respect to each DGP and estimator. 

\begin{figure}[!htbp]
    {\centering
    \caption{Results from Simulation: MBs reject $H_0$ when it is false and fail to reject when it is true.}
    \label{fig:simulation}
    \includegraphics[width=\textwidth]{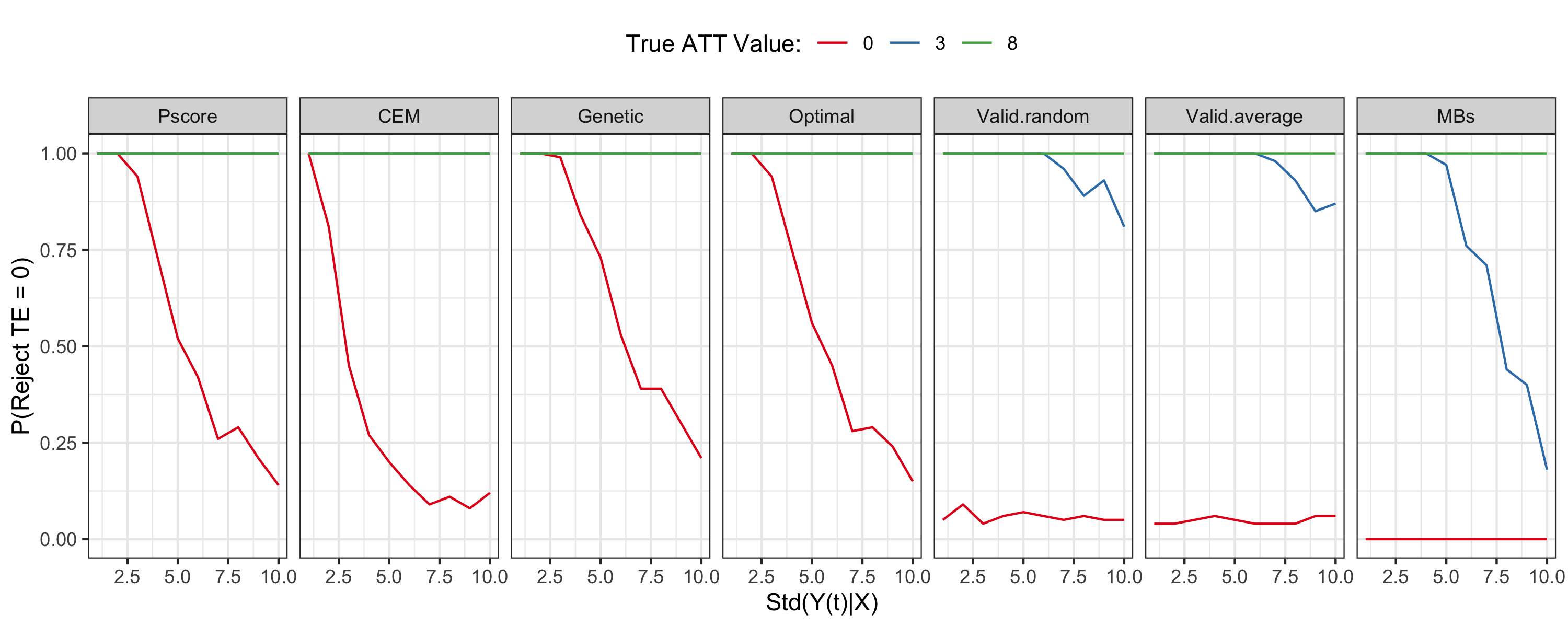}}
    \footnotesize{Note: Each point is the proportion of simulations in which $H_0: \tau = 0$ was rejected. Line color indicates value of the true ATT, $\tau$, and each panel represents a different matching method. Matching methods include: Propensity Score Matching (Pscore, left), Coarsened Exact Matching (CEM, second), Genetic Matching (Genetic, third), Optimal Matching (Optimal, fourth), Randomly chosen among valid assignments (Valid, fifth), the Average ATE among valid assignments (Valid average), and Matching Bounds (right). Standard errors for each estimate were computed using the sample standard error of each estimator, averaged over 100 simulations conducted at each point. The p-value for each test was computed with a standard normal distribution. Recall from the setup that  when $\tau=0$, we should not reject (i.e., the \textcolor{red}{red} curve should always be horizontal at zero), and when $\tau >0$ we should reject (\textcolor{blue}{blue} and \textcolor{green}{green} curves should be high). Note that MBs (rightmost algorithm) never rejects the null hypothesis when it it true. They are not too conservative when it is false.}
\end{figure}

Results from our simulations are reported in Figure \ref{fig:simulation}. At each value of $\sigma^2$ and $\tau$ (corresponding to each point on each plot), we performed 100 simulations and recorded the proportion of the 100 times that the null hypothesis $H_0: \tau = 0$ was rejected by each method we compared. We did this both in cases where the null hypothesis was true (red curves) and cases where it was false (all other curves). From the figure, we see that most popular matching methods tend to incorrectly reject $H_0$ when the true variance is low but the ATT is 0: \textit{this is because the estimates from other matching methods are far away from the true ATT}. We also see that \textit{MBs do not suffer from this problem: $H_0$ is never rejected when $\tau = 0$.} Additionally, we see that MBs \textit{do successfully reject} $H_0$ when $\tau\neq 0$.  When MBs is conservative, this is because the signal to noise ratio, i.e., the ratio between $\tau$ and $\sigma^2$, favors $\sigma^2$: the ATT is hard to detect among overly noisy data. Because of this, even the next best method (averaging among all minimal balance matched sets, denoted by Valid average in Figure \ref{fig:simulation}) fails to reject the null $\sim$6\% of the time when variance is large (right curve of Valid Average method in second to right plot). This $\sim$6\% false rejection rate is due solely from the matching procedure and (as we know) can be avoided using MBs.

\section{Applications}\label{sec:appl}
In this section we further test Matching Bounds by applying the methods to two examples of common questions in political science addressed by two studies which we replicate. With matching bounds we will be able to assess the robustness of both studies to the problem we  presented in this paper. We will see that the first study is largely robust to variation in matching assignment, while the second is less robust.

\subsection{Conflict-Induced Migration in Nepal}\label{sec:adi}
We apply the methods introduced in our paper to a study of the effects of conflict on migration conducted by~\cite{adhikari2013conflict}. We show that the findings of the study are robust to choice matched set by applying Matching Bounds. 

\cite{adhikari2013conflict} argues that individuals in areas experiencing conflict tend to migrate away if they directly experience violence. Their case study is Nepal, which experienced a civil war lasting longer than a decade starting in 1996. During this conflict, more than 50,000 were displaced. In 2008, \cite{adhikari2013conflict} conduct a survey of 1804 households in 11 districts of Nepal, plus the capital. The sample is stratified across several covariates to ensure that it is representative with respect to the variables of interest. 

The main treatment variable will be whether or not a surveyed individual experienced any violence prior to their decision or not to migrate. The outcome variable is also binary, coded as one if a responded was actually displaced during the conflict. The control covariates are as follows: age, sex, income, education level, number of children, and political affiliation of the respondent; as well as presence of working roads, industry and amount of land owned by the respondent in their initial place of residence. Finally, variables indicating whether industries, land, and homes were destroyed in the respondent's initial place of residence are also present.  

Due to the sampling framework separately targeting displaced and non-displaced respondents separately, \cite{adhikari2013conflict} conduct a robustness test of their main analyses using  propensity score-nearest neighbor matching with replacement.  They then estimate treatment effects on the matched data using linear regression. Here we will replicate this particular analysis, but we will vary which matched assignment is selected using our method. 

In order to mimic the analysis of the original paper, we also estimate the main effects of interest by employing outcome regression on each of the match assignments obtained with our method. The regression model also includes all of the matching variables as control covariates. The effect of interest is the Sample Average Treatment Effect on the Control (SATC): since there are many more treated than control units, we constrain each control unit to be matched to one treated unit. We construct matched groups by first employing nearest neighbor matching on the propensity score, which serves as the baseline original method used by \cite{adhikari2013conflict}, then we compute the following balance statistics on the matched set so obtained: difference in the first three moments of the distributions of each matched covariate, as well as absolute difference in propensity scores between each unit and its matched counterpart. We then constrain Matching Bounds to be within $\epsilon \times 100\%$ of these original values. This $\epsilon$ is to be understood as a level of tolerance for Matching Bounds: all matched groups whose balance statistics are within $\epsilon \times 100 \%$ of the respective statistics on the original matched sets will be admissable. We additionally assess balance by statistically testing whether the joint distribution of the covariates among the control units is different from its counterpart among the matched treatment units. We do so by employing the modified Cramer test of \cite{baringhaus2004new}. We note that this is a stricter criterion for balance than those most commonly employed, which tend to test only whether the marginal distributions of the covariates are statistically different before/after matching (e.g., \citealt{adhikari2013conflict} only tests for differences in means of the marginal distributions of the covariates).

\begin{figure}
    \centering
    \includegraphics[width=\textwidth]{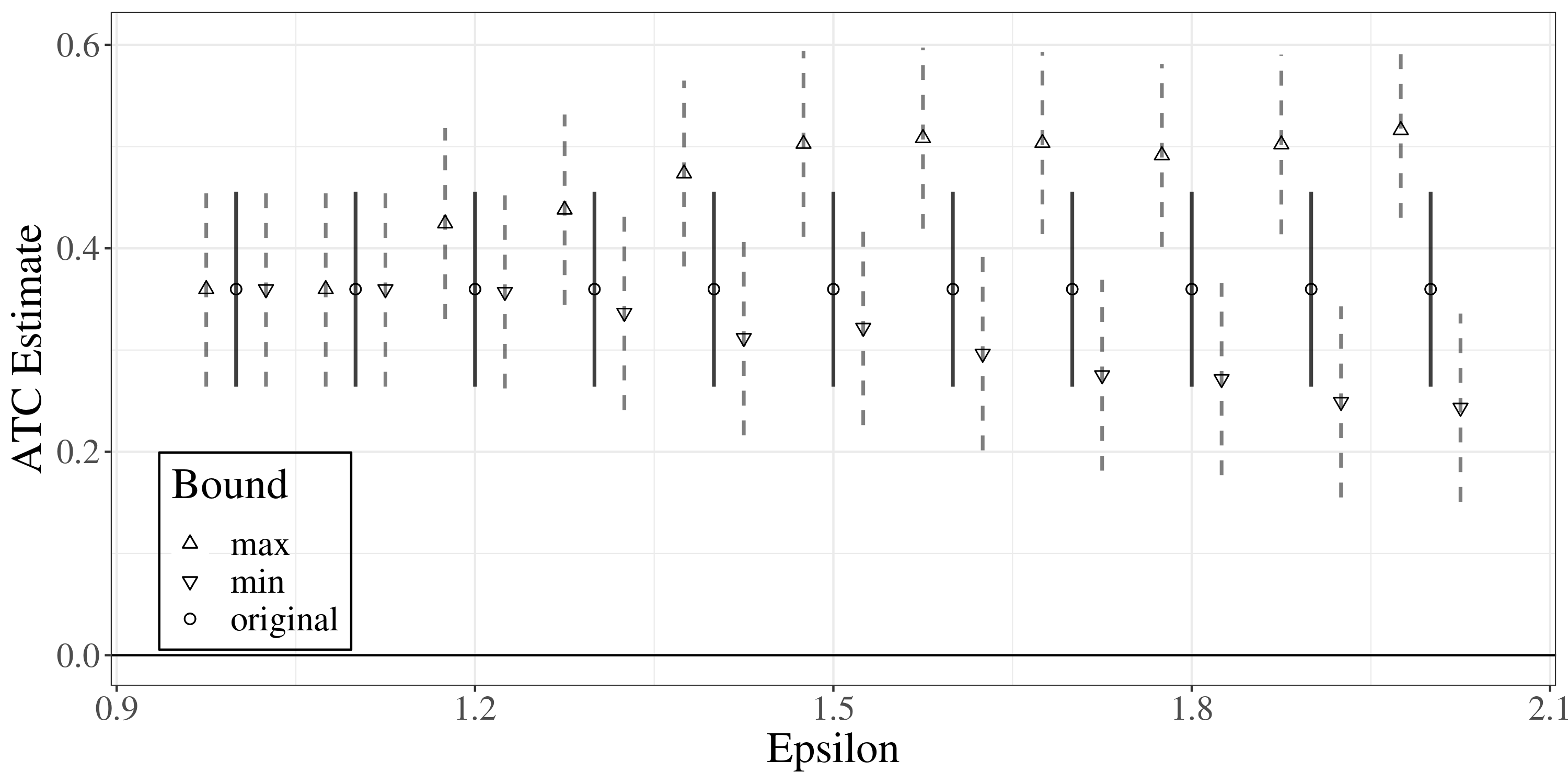}
    \includegraphics[width=\textwidth]{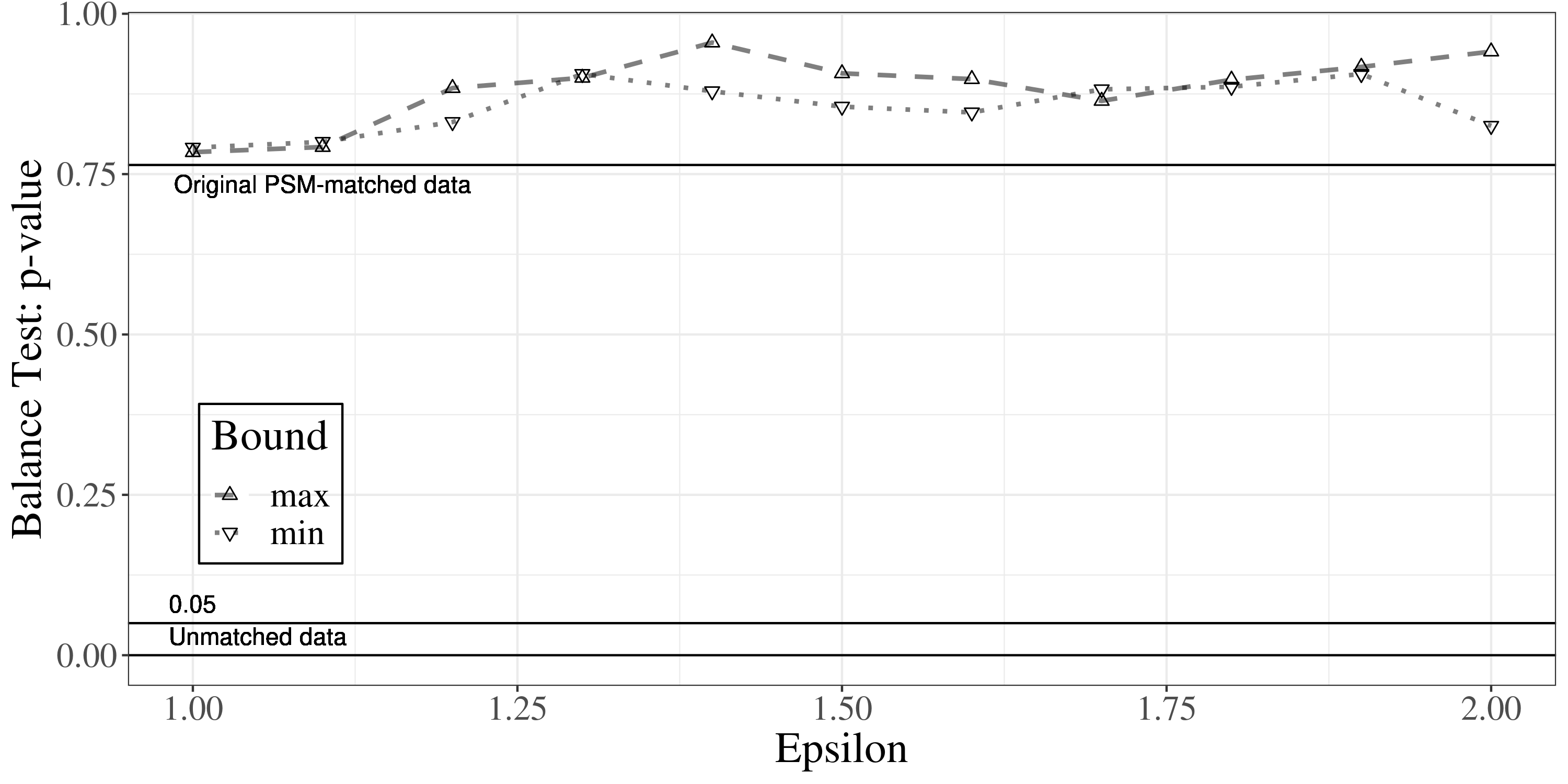}
    \caption{Results of Matching Bounds to the dataset of \cite{adhikari2013conflict}. Horizontal axis: proportion of allowed diversion from original matches. Top panel: ATC estimate using regression coefficient of treatment on outcome together with 95\% confidence intervals. Bottom panel: p-value of modified Cramer test to assess multivariate covariate balance. Higher values imply greater similarity between the joint distributions of the matched covariates in the treated and control groups. } 
    \label{fig:adires}
\end{figure}

Results are presented in Figure \ref{fig:adires}. The figure shows two clear conclusions: first even when we allow for matching bounds whose balance statistics are twice as far away as those of the original data, we still can identify treatment effects that are similar to those in the original matched data, and, importantly, on the same side of the zero line. Second, the bottom panel of Figure \ref{fig:adires} shows that the p-value for the joint balance test is well above the 0.05 line at all values of $\epsilon$ (note that the same test performed on the unmatched data is below this line). This implies that there is good reason to believe that an acceptable level of balance is achieved by the match assignments produced by Matching Bounds at all levels of $\epsilon$ we considered.

Our re-analysis of the data from \cite{adhikari2013conflict} shows that the original results are generally robust to the problem of multiple matched groups with similar balance.

\subsection{Partisan Shifts and Openness to Foreign Trade}
 We now apply Matching Bounds to a study, originally conducted by \cite{grieco2009preferences}, of countries' openness to foreign trade. The article studies (among other questions) whether countries with more left-leaning governments tend to impose more restrictions on international trade, even if these restrictions are in violation of existing international trade agreements they are part of.

The study sample is a panel of 182 countries measured yearly from 1975 to 2008. Among these, the sample is further reduced to only those countries that agreed to Article VIII of the International Monetary Fund's Articles of Agreement. This international treaty prohibits signatory countries from implementing restrictions to free trade with other signatories. The treatment variable is one if the government of the country is more left leaning than the government under which that country signed Article VIII. The outcome variable is a binary indicator for whether any trade restrictions were imposed by the country in a given year. Matching covariates are exchange rate flexibility, trade dependence, GNP per capita, GDP growth, reserves/GDP, reserves volatility, balance of payments/GDP, terms of trade volatility, IMF surveillance, use of IMF credits, universality of Article VIII, and regional trade restrictions. 

\cite{grieco2009preferences} perform their original analysis using Genetic Matching \citep{Diamond2013}, with the SATT as their target estimand, since there are more control units than treated in the sample, they then employ the Bias-Corrected matching estimator of \cite{Abadie2010} on the matched data to produce their final SATT estimates. In our replication, we will use Genetic Matching to obtain baseline matches, and then apply linear regression to the matched data in order to produce estimates of the SATT, including all the matching covariates as controls. This is done for comparability to the application to \cite{adhikari2013conflict}, however we note that results obtained with our methodology are not dissimilar to those obtained with the estimator originally employed by \cite{grieco2009preferences}.  As done in the application to \cite{adhikari2013conflict}, we measure balance in several different ways on the original matches obtained by Genetic Matching, and then constrain Matching Bounds to be within a certain level of those original measures. Specifically, we use the difference in the first 3 moments of the marginal distributions of each covariate, as well as a caliper on the distance (where we use the distance  function produced by Genetic Matching) between each treated unit and its control match. Since the target estimand is the ATT, Matching Bounds are constrained to match each treated unit to one control. As done in Section \ref{sec:adi}, we will compute Matching Bounds at different levels of allowed distance (denoted by $\epsilon$) from the original levels of balance achieved by Genetic Matching.

\begin{figure}
    \centering
    \includegraphics[width=\textwidth]{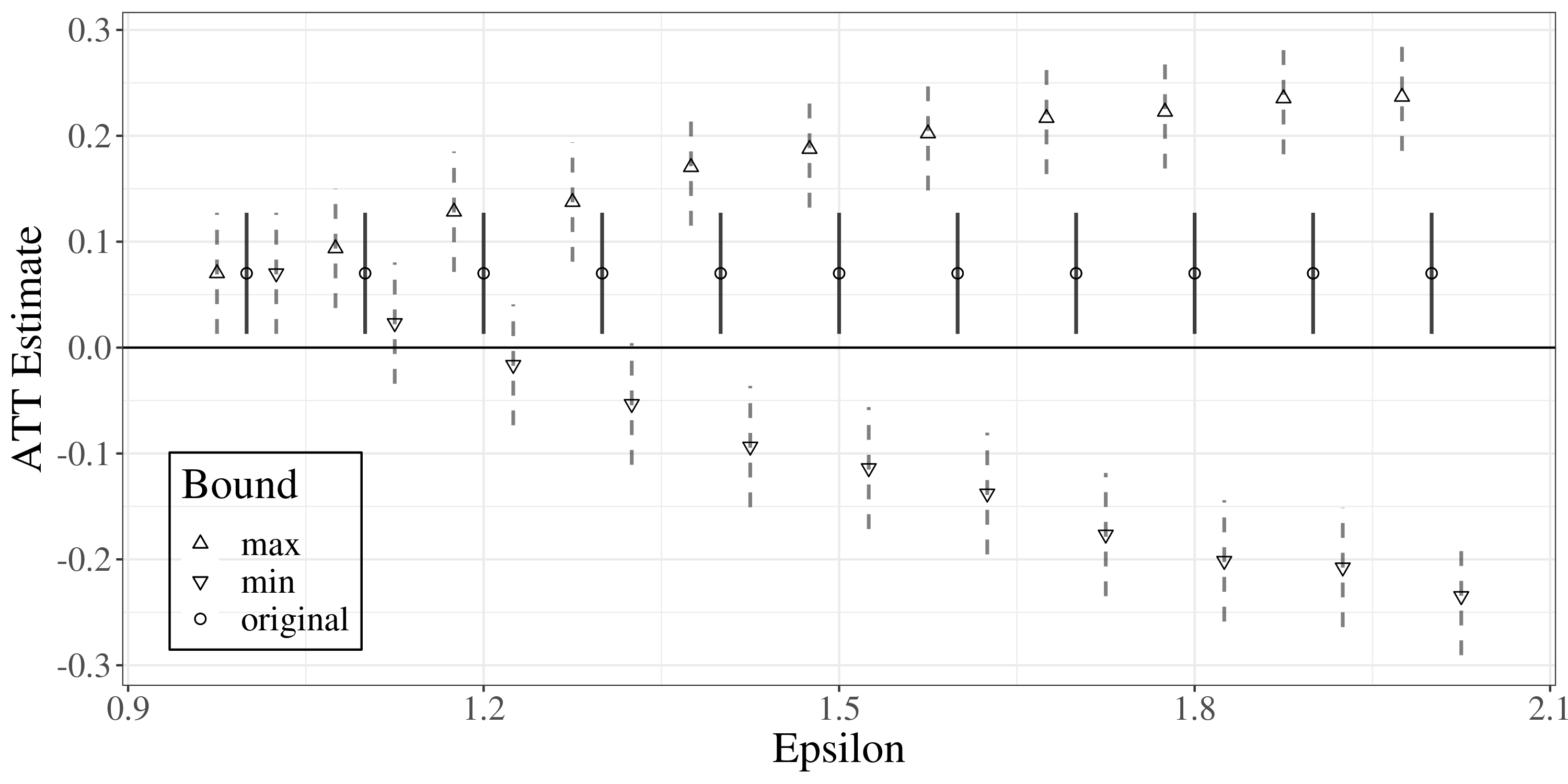}
    \includegraphics[width=\textwidth]{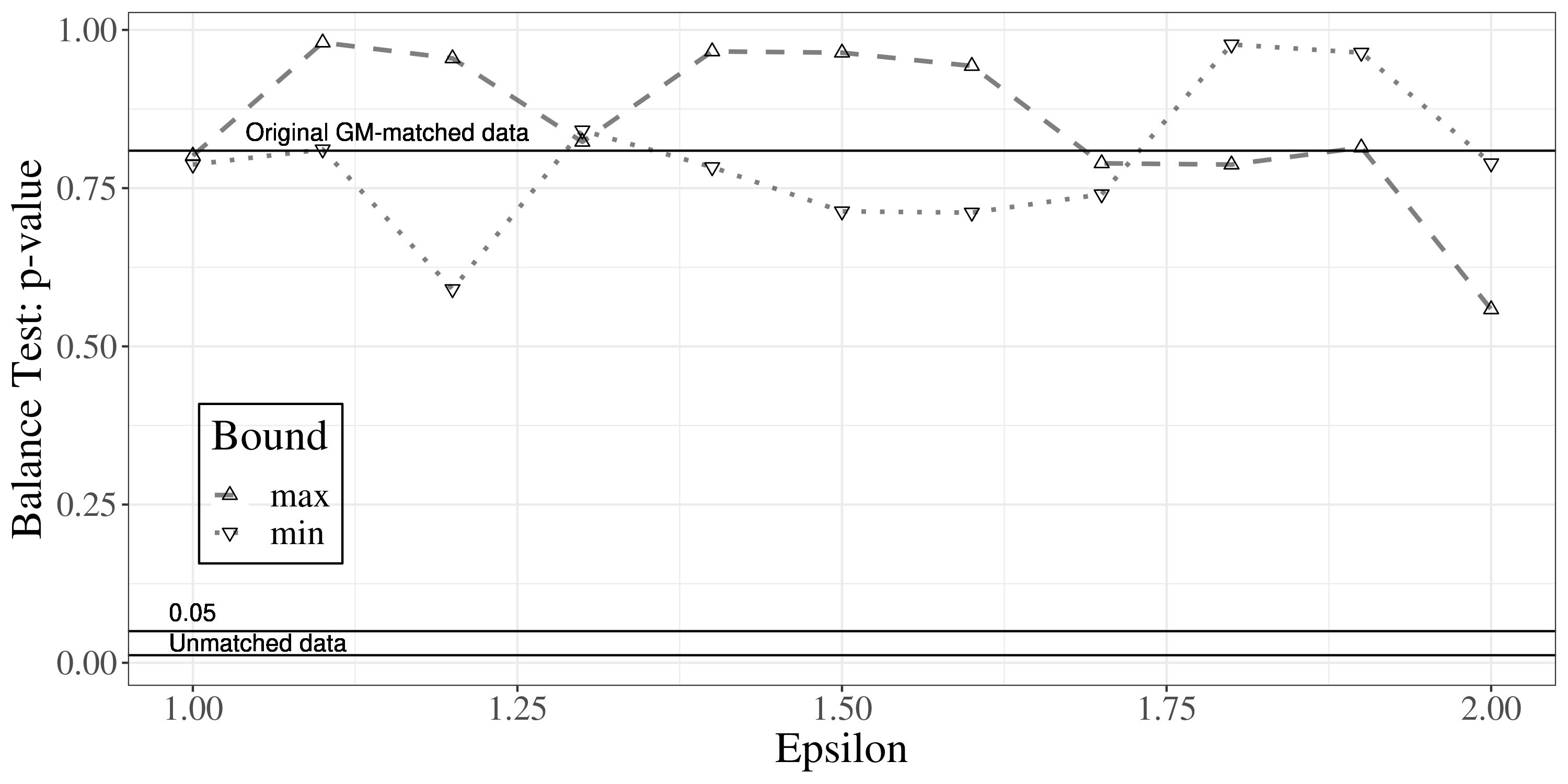}
    \caption{Results of Matching Bounds to the dataset of \cite{grieco2009preferences}. Horizontal axis: proportion of allowed diversion from original matches.  Top panel: ATT estimate using regression coefficient of treatment on outcome. Bottom panel: p-value of modified Cramer test to assess multivariate covariate balance. Higher values imply greater similarity between the joint distributions of the matched covariates in the treated and control groups. }
    \label{fig:warres}
\end{figure}

Results are presented in Figure \ref{fig:warres}: the top panel shows ATT effect estimates at increasingly more tolerant balance constraints.  Clearly, even a small relaxation of the balance criteria is sufficient to obtain bounds that cross the 0 line. In addition, the bottom panel of Figure \ref{fig:warres} shows balance achieved at each bound as measured by the p-value of the modified Cramer test of \cite{baringhaus2004new} described in Section \ref{sec:adi}: this value is well above the 0.05 line of rejection for all the bounds considered, and generally similar to that achieved by the original matched data. This is evidence that an acceptable level of balance is achieved at all levels of $\epsilon$, even though  effect estimates are much different from each other. 

Overall, our replication shows that there indeed exist matched sets that achieve similar balance, but lead to different answers on this dataset. This implies that either different techniques or more data might be needed in order to address the question of the original study.

\section{Conclusion}
In this paper we have argued that matching algorithms that perform equally well in terms of existing metrics can lead to vastly different treatment effects estimates because  multiple  matches of similar quality can exist for one unit. We have introduced Matching Bounds, a method to account for the uncertainty stemming from choice of matched sample. Matching Bounds uses robust optimization, and allows analysts to produce results that are robust to the arbitrary choice of match assignment.

We wish to clarify that the extra source of uncertainty we have studied here should not be used as a reason to avoid matching methods:
matching is a fundamental tool for observational inference with several desirable properties, such as being powerful and nonparametric, as well as easily interpretable. We know of no other current method for causal inference with both of these properties.
Regression adjustment is not a viable alternative to matching because a) it usually imposes strong parametric assumptions on the outcome model, and b) existing results \citep{Angrist2008, imai2016, Aronow2016}  have shown that regression is in itself a form of weighting, which is itself close to matching, and as such it is not exempt from any of the problems outlined in this paper.

In an age in which empirical evaluations inform scientific theory and policy more and more, it is fundamental that empirical results are correct and robust.
In that sense, implications of this study for applied fields such as political economy and development economics are vast: simply, by using Matching Bounds, as opposed to a arbitrary matching method, \textit{results based on data are less likely to be wrong}.
With Matching Bounds, it will be possible to obtain inferences that are robust to matching method choice, thus providing stronger empirical grounds for both theory evaluation and policy-making.  
\clearpage

\bibliography{biblio}

\clearpage

\appendix
\begin{center}
\LARGE{Matching Bounds: How Choice of Matching Algorithm Impacts Treatment Effects Estimates and What to Do about It.}\\
\LARGE{\textbf{Supplement}}
\end{center}
\thispagestyle{empty}
\setcounter{page}{1}
\setcounter{figure}{0}    
\setcounter{table}{0}    

\etocdepthtag.toc{mtappendix}
\etocsettagdepth{mtchapter}{none}
\etocsettagdepth{mtappendix}{subsection}
\tableofcontents
\clearpage

\section{Derivations in Section \ref{sec:howandwhy}}
Here we give a detailed derivation of the quantities in Section \ref{sec:howandwhy}. We start with a derivation of the decomposition in \eqref{Eq:LinApprox}. Recall that we have defined $Y(t) = f_t(\bx, \bu, \epsilon)$ with $\E[\epsilon] = 0$. Let $N_a^c = \sum_{i=1}^{N^t}\sum_{j=1}^{N^c}w_{ij}^a$ be the sum of all matches in group $a$, and let $N^b_c$ be the same quantity for match assignment $b$. All quantities are as defined in the main paper. 
\begin{align*}
\tauhat^a - \tauhat^b &= \frac{1}{N_c^a}\sum_{i = 1}^NY_i(0)(1-T_i)w_i^a - \frac{1}{N_c^b}\sum_{i = 1}^NY_i(0)(1-T_i)w_i^b\\
&= \frac{1}{N_c^a}\sum_{i = 1}^Nf_0(\bx_i, \bu_i, \epsilon_i)(1-T_i)w_i^a - \frac{1}{N_c^b}\sum_{i = 1}^Nf_0(\bx_i, \bu_i, \epsilon_i)(1-T_i)w_i^b.
\end{align*}
Now take a Taylor expansion of $f_0(\bx, \bu, \epsilon)$ at $(\mathbf{0}, \mathbf{0}, 0)$. Let $\nabla f_0(\mathbf{0}, \mathbf{0}, 0) = (\boldsymbol{\beta}_1, \boldsymbol{\beta}_2, \beta_3)^T$, and let $f_0(\mathbf{0}, \mathbf{0}, 0) = \alpha$. We have:
\begin{align*}
f_0(\bx, \bu, \epsilon) &= \alpha + \boldsymbol{\beta_1}^T\bx + \boldsymbol{\beta_2}^T\bu + \beta_3\epsilon + \text{higher order terms}.
\end{align*}
If we substitute this into the difference we started with, we obtain:
\begin{align*}
\tauhat^a - \tauhat^b &= \frac{1}{N_c^a}\sum_{i = 1}^N(\alpha + \boldsymbol{\beta_1}^T\bx_i + \boldsymbol{\beta_2}^T\bu_i + \beta_3\epsilon_i)(1-T_i)w_i^a \\
&- \frac{1}{N_c^b}\sum_{i = 1}^N(\alpha + \boldsymbol{\beta_1}^T\bx_i + \boldsymbol{\beta_2}^T\bu_i + \beta_3\epsilon_i)(1-T_i)w_i^b\\
&+ \text{difference of higher order terms}.
\end{align*}
For simplicity, now define $\bar{\bx}^a = \frac{1}{N_c^a}\sum_{i=1}^N\bx_i(1-T_i)w_i^a$, the average observed covariate values under match assignment $\W^a$,  $\bar{\bu}^a = \frac{1}{N_c^a}\sum_{i=1}^N\bu_i(1-T_i)w_i^a$, the average unobserved covariates value under $\W^a$, and $\bar{\epsilon}^a = \frac{1}{N_c^a}\sum_{i=1}^N\epsilon_i(1-T_i)w_i^a$, the average noise under $\W^a$. Let $\bar{\bx}^b, \bar{\bu}^b, \bar{\epsilon}^b$ be defined analogously for match assignment $\W^b$. We can plug this into the above to obtain:
\begin{align}
\tauhat^a - \tauhat^b &= \underbrace{\boldsymbol{\beta}_1^T(\bar{\bx}^a - \bar{\bx}^b)}_{\text{Observables}} + \underbrace{\boldsymbol{\beta}_2^T(\bar{\bu}^a - \bar{\bu}^b)}_\text{Unobservables} + \underbrace{\beta_3(\bar{\epsilon}^a - \bar{\epsilon}^b)}_\text{Conditional Variance}\nonumber\\
&+ \text{difference of higher order terms}.\nonumber
\end{align}
To derive the upper bound on $|\bar{\epsilon}^a - \bar{\epsilon}^b|$ in Section \ref{sec:why}, we can apply the Cauchy-Schwartz inequality to the quantity as follows: 
\begin{align*}
\E[|\bar{\epsilon}^a - \bar{\epsilon}^b|] &\leq \sqrt{\E\left[(\bar{\epsilon}^a - \bar{\epsilon}^b)^2\right]}\\
&= \sqrt{\E[(\bar{\epsilon}^a)^2] - 2\E[\bar{\epsilon}^a\bar{\epsilon}^b] + \E[(\bar{\epsilon}^b)^2]}\\
&= \sqrt{Var(\bar{\epsilon}^a) + Var(\bar{\epsilon}^b)}\\
&= \sqrt{Var(\bar{Y}^a|\bX, \bU) + Var(\bar{Y}^b|\bX, \bU)},
\end{align*}
where the inequality comes from the Cauchy-Schwartz inequality, first equality from expanding the square, the second from the definition of variance and the fact that $\E[\bar{\epsilon}^a] = \E[\bar{\epsilon}^b] = 0$ by assumption, and that $\bar{\epsilon}^a$ and $\bar{\epsilon}^b$ are independent for non-random match assignments $\mathbf{W}^a, \mathbf{W}^b$, as is the case here. The third equality comes from the definition of $Y(0)$.

\section{AJPS Replication Methodology}\label{Appendix:Replication}
Results are replicated using the original author's covariate set for matching and model specification for regression. Since virtually all of the papers replicated in this study make use of matching in conjunction with regression \citep{Ho2007}, we do the same here. We try to remain as close to the original author's preprocessing and post-matching estimation as possible, changing only which method is used to construct the matches. This means that we run the same regression models that the authors do after matching. Each paper produces results using different combinations of either matching covariates, treatment and dependent variables, or post-matching regression adjustment. We replicate as many of the combinations used by the original authors as possible. Each row in Figures \ref{Res1} and \ref{Bal1} and in Tables \ref{A:Res1} and \ref{A:Bal1} is one such combination. This explains why there are many different results for a single paper.

The original results reported by the authors as well as paper titles and references are willingly omitted from this replication, as the results reported here are not meant to directly question the findings of the original authors, but, rather to analyze the problem of matching dependency of results stemming from real-life data and models. Indeed, the authors might have perfectly good reasons to prefer the matching method they choose to others and to believe that the results they obtain with that method are the correct ones. The analysis here should be viewed not much as a replication but more as a simulation on real-life data, where the data belongs to actual scientific studies and the hypotheses stem from actual theories, rather than computer simulations.

Table~\ref{Emp:T1}  presents a summary of the algorithms used in this replication.\footnote{All the algorithms employed here use the same loss: the sum of the Mahalanobis distances between matched units with no calipers or balance constraints.} These specific algorithms were chosen because they are among those used in the replicated paper and because, in general, they have been quite popular in applied research across all fields of political science.

\begin{table}[h]
\resizebox{\textwidth}{!}{%
\begin{tabular}{rrrr}
\hline
Code& Citaton& Method & Target \\
\hline
No Matching& & OLS regression & Approximate minimization of L2 distance between pairs.\\
CEM& \citet{Iacus2012}& Coarsened Exact Matching& Upper bound on distance between units.\\
GenMatch&\citet{Diamond2013}& Genetic Matching& Minimization of loss.\\
OptMatch&\citet{Rosenbaum1989}& Optimal Matching& Minimization of loss.\\
Nearest& &Nearest Neighbor& Approximate minimization of loss.\\
MIP-Match&\citet{Zubizarreta2012}& Mixed Integer Matching&Minimization of loss.\\
\hline
\end{tabular}}
\caption{Matching methods employed in this replication.}
\label{Emp:T1}
\end{table}
The results from the matching were used as weights in the subsequent regressions: units with a total weight of 0 were dropped from the sample, while units with weights greater than 1 were replicated as many times as they were matched. We do not explicitly employ the difference-in-means estimator that has been studied in the theoretical section of this paper because we prioritize fidelity to the original authors methodology and model specification. Replication failed in a few instances: this could be due either  either matching problems not being solvable given covariate specifications or regression models not being estimable. We chose to report failure in these cases, instead of altering the pool of matching covariates or the statistical model we chose to estimate.

Balance results were computed using the Cobalt R package. We used standardized absolute difference between the mean of each covariate in the treated and control group respectively. Standardization was conducted using the standard deviation of the whole unweighted sample, as this is a more conservative estimate of matched standard deviation. If this measure of balance is below a threshold of 0.1, then we consider that covariate balanced, as recommended by \citet{Stuart2013}. Figure \ref{Bal1} and Table \ref{A:Bal1} report percentages of covariates balanced by each method.

The dataset used by the original authors in paper number 10 was too large to be used with some of the matching methods tested (GenMatch, OptMatch, Mip-Match). For this reason, we took a random sample of the observations (N=2000, roughly 10\% of the original datasets) from the data to perform our tests.
\subsection{Numeric versions of Figures \ref{Res1} and \ref{Bal1}}
\begin{sidewaystable}[!htbp]
\scriptsize
\centering
\begin{tabular}{rrrrrrr|rrrrrrr}
  \hline
Paper & No Matching & CEM & GenMatch & OptMatch & Nearest & MIP-Match & Paper & No Matching & CEM & GenMatch & OptMatch & Nearest & MIP-Match \\
  \hline
    1 & 8.83 & 7.68 & 6.79 & 5.35 & 6.16 & 3.09      &  8 & 2.26 &  & 2.41 & 1.69 & 2.66 & 2.31 \\
    1 & 8.88 & 8.55 & 7.75 & 5.49 & 6.14 & 3.10      &  8 & 0.02 &  & -1.23 & -0.27 & 0.87 & 0.01\\
    2 & 0.18 &      & -0.83 & -0.13 & 0.08 & 0.29    &  8 & -1.93 &  & -1.88 & -1.13 & -1.20 &-1.86 \\
    3 & 1.80 & 1.32 & -0.05 & 1.86 & 0.05 & 0.00     &  8 & 0.46 &  & 1.44 & 0.34 & 0.86 & 0.68 \\
    3 & 1.36 & 2.89 & -0.02 & 2.60 & 1.80 & 1.32     &  8 &  &  &  &  &  &  \\
    3 & -1.53 & 0.04 & -0.24 & 1.10 & 1.36 & 2.89    &  8 &  &  &  &  &  &  \\
    4 & 2.35 & 0.00 & 1.20 & 1.34 & 1.37 & 2.22      &  8 &  &  &  &  &  &  \\
    4 & 2.38 & -0.00 & 1.00 & 1.65 & 1.66 & 2.06     &  8 &  &  &  &  &  &  \\
    4 & 1.80 & -0.05 & 1.08 & 0.81 & 0.94 & 0.79     &  9 & -2.18 &  & 1.29 & -5.41 & -1.32 &  \\
    4 & 2.02 & 1.34 & 1.14 & 1.03 & 1.21 & 3.12      &  9 & -1.32 &  & -2.16 & 2.01 & -5.51 & 0.40 \\
    5 & 16.52 & -1.61 &  &  & 1.78 &                 &  9 & -5.51 & 0.40 & -2.11 & -1.24 & 0.52 & 0.92 \\
    5 & 17.86 &  & 9.79 & 24.81 &  & 1.39            &  9 & 0.52 & 0.92 & 0.48 & -2.09 & -2.40 & -0.43 \\
    5 & 17.87 &  &  & 27.00 & 16.52 & -1.61          &  9 & -2.40 & -0.43 & -1.76 & 0.25 & -2.15 & -4.35 \\
    5 &  & 5.46 &  & 22.94 & 17.86 &                 &  9 & -2.15 & -4.35 & -2.28 & -2.72 & 0.58 & -0.83 \\
    5 &  &  & 9.73 &  & 17.87 &                      &  9 & 0.58 & -0.83 & -1.57 & -2.74 & -2.32 & -4.13 \\
    6 & -1.03 & -2.72 & -1.54 & -1.45 & 0.25 &       &  9 & -2.32 & -4.13 & -0.86 & -1.49 & -2.27 & -0.24 \\
    6 & 0.39 & -0.29 & -0.09 & -0.29 & -2.07 & 2.46  &  9 & -2.27 & -0.24 & -5.42 & 0.90 &  & -1.43 \\
    6 & -2.12 & 0.58 & -1.27 & -0.63 & -1.03 & -2.72 &  9 &  & -1.43 & 1.04 & -5.41 &  & 0.13 \\
    6 &  & -1.42 & -1.58 & -1.63 & 0.39 & -0.29      &  9 &  & 0.13 & -2.27 & 1.07 &  & 1.77 \\
    6 &  & -0.42 & 0.20 & -0.96 & -2.12 & 0.58       &  9 &  & 1.77 & -2.19 & 0.37 &  & -1.52 \\
    6 & 1.59 & -2.03 & 1.58 & 0.49 & -2.53 &         &  9 &  & -1.52 & 0.42 & -2.23 &  & -3.14 \\
    6 & -2.69 & 1.98 & -2.63 & 1.03 & 0.57 & 2.09    &  9 &  & -3.14 & -1.86 & 0.73 &  & -1.62 \\
    6 & 0.46 & 1.55 & 0.49 & -2.35 & 1.59 & -2.03    &  9 &  & -1.62 & -2.29 & -0.85 &  & -1.24 \\
    6 &  & -2.25 & 1.58 & 1.29 & -2.69 & 1.98        &  9 &  & -1.24 & -1.86 & -2.18 &  & -5.49 \\
    6 &  & 1.20 & -2.63 & 1.78 & 0.46 & 1.55         &  9 &  & -5.49 & -0.56 & -2.18 &  & 1.29 \\
    6 & 0.16 & -2.56 & -0.08 & 0.00 & -1.33 &        &  10 & 0.04 & -0.32 & -0.54 & 0.33 & 0.31 & -0.09 \\
    6 & -0.57 &  & -1.17 & -0.76 & 0.62 & 14.76      &  10 & 0.17 & -0.16 & -0.68 & 0.33 & 0.41 & 0.04 \\
    6 & 0.24 & -0.06 & 0.18 & 0.15 & 0.16 & -2.56    &  10 & 0.04 & -0.32 & -0.68 & 0.42 & 0.31 & 0.17 \\
    6 &  & 1.95 & -0.22 & 0.76 & -0.57 &             &  10 & 0.04 & -0.32 & -0.54 & 0.33 & 0.29 & 0.04 \\
    6 &  & -0.18 & -0.84 & 0.24 & 0.24 & -0.06       &  10 & 0.17 & -0.16 & -0.68 & 0.31 & 0.41 & 0.04 \\
    6 & 1.58 & -2.41 &  &  & -2.93 &                 &  10 & 0.04 & -0.32 & 0.33 & 0.41 & 0.29 & 0.17 \\
    6 & -3.24 & 1.05 &  & 1.28 & 0.53 &              &  10 &  & -0.68 & 0.42 & 0.31 & 0.29 & 0.04 \\
    6 & 1.01 & -1.73 &  & -2.69 & 1.23 & &&&&&&& \\
    6 &  & -3.75 &  & 0.46 & -2.81 & &&&&&&& \\
    6 &  & 0.25 &  & 3.06 & 0.98 & 0.29 &&&&&&& \\
    7 & 9.07 &  & 3.77 & 3.71 & 5.21 & 7.36 &&&&&&& \\
   \hline
\end{tabular}
\caption{Results of replicating 10 main results of AJPS papers with different matching methods. Each row is a hypothesis test conducted on a different combination of matching, treatment, outcome variables or post-matching estimator. Every one of these combinations was used in the original paper. Each entry is the value of a Z statistic: values below -1.96 and above 1.96 can be interpreted as statistically significant at the conventional threshold. Blank entries correspond to instances in which either the matching method or the post-matching estimator failed to run on the specific combination of variables. The table shows that different matching methods produce different TE estimates on the same datasets, even when they induce similar balance.  }
\label{A:Res1}
\end{sidewaystable}

\begin{sidewaystable}[!htbp]
\scriptsize
\centering
\begin{tabular}{rrrrrrr|rrrrrrr}
  \hline
Paper & No Matching & CEM & GenMatch & OptMatch & Nearest & MIP-Match & Paper & No Matching & CEM & GenMatch & OptMatch & Nearest & MIP-Match \\
  \hline
    1 & 1.00 & 1.00 & 1.00 & 0.80 & 0.80 & 0.60   &      8 & 1.00 & 1.00 & 1.00 & 1.00 & 1.00 & 1.00 \\
    1 & 1.00 & 1.00 & 1.00 & 0.80 & 0.80 & 0.60   &      8 & 1.00 & 1.00 & 1.00 & 1.00 & 1.00 & 1.00 \\
    2 & 0.14 & 0.71 & 0.43 & 0.14 & 0.14 & 0.29   &      8 & 1.00 & 1.00 & 1.00 & 1.00 & 1.00 & 1.00 \\
    3 & 0.97 & 1.00 & 1.00 & 1.00 & 1.00 & 1.00   &      8 & 1.00 & 1.00 & 1.00 & 1.00 & 1.00 & 1.00 \\
    3 & 0.97 & 1.00 & 1.00 & 1.00 & 1.00 & 1.00   &      8 & 1.00 & 1.00 & 1.00 & 1.00 & 1.00 & 1.00 \\
    3 & 0.97 & 1.00 & 1.00 & 1.00 & 1.00 & 1.00   &      8 & 1.00 & 1.00 & 1.00 & 1.00 & 1.00 & 1.00 \\
    4 & 0.25 & 1.00 & 1.00 & 0.25 & 0.25 & 0.50   &      8 & 1.00 & 1.00 & 1.00 & 1.00 & 1.00 & 0.00 \\
    4 & 0.25 & 1.00 & 1.00 & 0.25 & 0.25 & 0.50   &      8 & 1.00 & 1.00 & 1.00 & 1.00 & 1.00 & 0.00 \\
    4 & 0.25 & 1.00 & 1.00 & 0.25 & 0.25 & 0.50   &      9 & 0.80 & 1.00 & 1.00 & 0.80 & 0.80 & 0.80 \\
    4 & 0.25 & 1.00 & 1.00 & 0.25 & 0.25 & 0.50   &      9 & 0.80 & 1.00 & 1.00 & 0.80 & 0.80 & 0.80 \\
    5 & 0.50 & 0.67 & 0.33 & 0.17 & 0.17 & 0.00   &      9 & 0.80 & 1.00 & 1.00 & 0.70 & 0.80 & 0.80 \\
    5 & 0.33 & 0.67 & 0.17 & 0.00 & 0.00 & 0.00   &      9 & 0.80 & 1.00 & 1.00 & 0.70 & 0.80 & 0.80 \\
    5 & 0.33 & 0.83 & 0.50 & 0.33 & 0.17 & 0.00   &      9 & 0.80 & 1.00 & 1.00 & 0.80 & 0.80 & 0.80 \\
    5 & 0.50 & 0.67 & 0.33 & 0.17 & 0.17 & 0.00   &      9 & 0.80 & 1.00 & 1.00 & 0.80 & 0.80 & 0.80 \\
    5 & 0.33 & 0.67 & 0.17 & 0.00 & 0.00 & 0.00   &      9 & 0.81 &  & 0.85 & 0.77 & 0.77 & 0.73 \\
    6 & 1.00 & 1.00 & 1.00 & 0.80 & 0.80 & 1.00   &      9 & 0.81 &  & 0.85 & 0.77 & 0.77 & 0.73 \\
    6 & 1.00 & 1.00 & 1.00 & 0.80 & 0.80 & 1.00   &      9 & 0.69 &  & 0.77 & 0.58 & 0.58 & 0.69 \\
    6 & 1.00 & 1.00 & 1.00 & 0.80 & 0.80 & 1.00   &      9 & 0.69 &  & 0.77 & 0.58 & 0.58 & 0.69 \\
    6 & 1.00 & 1.00 & 1.00 & 0.80 & 0.80 & 1.00   &      9 & 0.77 &  & 0.88 & 0.77 & 0.77 & 0.77 \\
    6 & 1.00 & 1.00 & 1.00 & 0.80 & 0.80 & 1.00   &      9 & 0.77 &  & 0.88 & 0.77 & 0.77 & 0.77 \\
    6 & 1.00 & 1.00 & 1.00 & 0.80 & 0.80 & 1.00   &      9 & 0.81 & 0.78 & 0.83 & 0.81 & 0.81 & 0.78 \\
    6 & 0.80 & 1.00 & 1.00 & 0.80 & 0.80 & 1.00   &      9 & 0.81 & 0.78 & 0.83 & 0.81 & 0.81 & 0.78 \\
    6 & 0.80 & 1.00 & 1.00 & 0.80 & 0.80 & 1.00   &      9 & 0.72 & 0.81 & 0.78 & 0.69 & 0.72 & 0.72 \\
    6 & 0.80 & 1.00 & 1.00 & 0.80 & 0.80 & 1.00   &      9 & 0.72 & 0.81 & 0.78 & 0.69 & 0.72 & 0.72 \\
    6 & 0.80 & 1.00 & 1.00 & 0.80 & 0.80 & 1.00   &      9 & 0.78 & 0.81 & 0.83 & 0.78 & 0.78 & 0.78 \\
    6 & 0.60 & 1.00 & 1.00 & 0.80 & 0.80 & 0.40   &      10 & 0.20 & 0.80 & 0.80 & 0.20 & 0.20 & 0.20 \\
    6 & 0.60 & 1.00 & 1.00 & 0.80 & 0.80 & 0.40   &      10 & 0.20 & 0.80 & 0.80 & 0.20 & 0.20 & 0.20 \\
    6 & 0.60 & 1.00 & 1.00 & 0.80 & 0.80 & 0.40   &      10 & 0.20 & 0.80 & 0.80 & 0.20 & 0.20 & 0.20 \\
    6 & 0.60 & 1.00 & 1.00 & 0.80 & 0.80 & 0.40   &      10 & 0.17 & 0.67 & 0.83 & 0.17 & 0.17 & 0.17 \\
    6 & 0.60 & 1.00 & 1.00 & 0.80 & 0.80 & 0.40   &      10 & 0.17 & 0.67 & 0.83 & 0.17 & 0.17 & 0.17 \\
    6 & 1.00 & 1.00 & 1.00 &  & 1.00 & 1.00   &          10 & 0.17 & 0.67 & 0.83 & 0.17 & 0.17 & 0.17 \\
    6 & 1.00 & 1.00 & 1.00 &  & 1.00 & 1.00   &          10 & 0.20 & 0.80 & 0.80 & 0.20 & 0.20 & 0.20 \\
    6 & 1.00 & 1.00 & 1.00 &  & 1.00 & 1.00 \\
    6 & 1.00 & 1.00 & 1.00 &  & 1.00 & 1.00 \\
    6 & 1.00 & 1.00 & 1.00 &  & 1.00 & 1.00 \\
    7 & 0.33 & 1.00 & 1.00 & 0.17 & 0.33 & 0.17 \\
   \hline
\end{tabular}
\caption{Results of replicating 10 AJPS papers with different matching methods. Each row is a hypothesis test conducted on a different combination of matching, treatment, outcome variables or post-matching estimator. Every one of these combinations was used in the original paper. Each entry is the proportion of covariates that were balanced after matching. A covariate is considered balanced if its standardized absolute difference in means between treated and control units is below 0.1. Blank entries correspond to instances in which the matching method failed to run on the specific combination of variables}
\label{A:Bal1}
\end{sidewaystable}

\newpage
\normalsize

\section{Constraining the Quality of Matches.}\label{Sec:Constraints}
Before Matching Bounds can be computed, it is necessary to define $\mathcal{C}$, the set of quality constraints that matches should satisfy to be acceptable. This will, in turn, restrict the space of possible matches we can choose from to the set $\OCD$. As argued before, there is no general proven set of criteria that matches should satisfy to be good, and therefore it is recommended that constraints be chosen on a case-by-case basis. Nonetheless, there have been several important contributions recommending criteria to determine whether the matches made are good, chiefly balance~\citep{rubin2006,Rosenbaum2007,Stuart2013} and aggregate distance between matched units~\citep{Diamond2013}. In this section we give constraint formulations for some of these popular metrics for both estimands and estimators introduced above

Starting with aggregate distance, several matching methods work by explicitly minimizing the sum of distances between matched pairs~\citep{Rosenbaum1989, Diamond2013, Zubizarreta2012}. The distance between any two units, $d_{ij}$ is defined as some metric on the covariate space. We might be interested in finding bounds around matches whose aggregate distance is below some constant $L$, perhaps one generated by first running any of the matching algorithms that directly optimize distance, and then using the $L$ arising from that algorithm. In this case we should constrain the matches as follows:
\begin{subequations}
\begin{align}
&\mbox{in Formulation 1: } && \sum_{i=1}^{N^t}\sum_{j=1}^{N^c} d_{ij}w_{ij} \leq L\\
&\mbox{in Formulation 2: } && \sum_{i=1}^{N^t}\sum_{j=1}^{N^c} d_{ij}w_{ij} \leq L\\
&\mbox{in Formulation 3: } && \sum_{i=1}^{N^t}\sum_{j=1}^{N^c} d_{ij}u_{ij} \leq L\cdot z.
\end{align}
\end{subequations}
We might also be interested in imposing a caliper, $C$, on the distance between two observations, an upper bound on the maximum distance for which two units are permitted to be matched. In this case it is easiest to preprocess the data to generate a matrix of auxiliary variables $D$ in which each entry is either 1 or 0 depending on whether the distance between two units is below the caliper size, and then by requiring $w_{ij} \leq D_{ij}$ for Formulation 1 and 2, and $u_{ij} \leq D_{ij}z$ for Formulation 3.

Balance is defined as similarity between the empirical distributions of the covariates in the treated and control samples~\citep{Imai2008}; there have been several suggestions as to how balance should be assessed and properly measured. Here we give constraint formulations for assessing similarity between moments of the covariate distributions as well as quantiles. Starting with moments, constraining the $k^{th}$ moment of the two distributions of the $p^{th}$ covariate to be almost $\sigma_p^k$ apart can be done by requiring:
\begin{subequations}
\begin{align}
&\mbox{in Formulation 1: } &&\biggl|\frac{1}{N^t}\sum_{i=1}^{N^t}(x_{ip}^t)^k- \frac{1}{N^t}\sum_{i=1}^{N^t}\sum_{j=1}^{N^c}w_{ij}(x^c_{jp})^k \biggr|\leq \sigma_p^k\label{Eq:Bal:Mom:F4}\\
&\mbox{in Formulation 2: } &&\biggl| \frac{1}{M}\sum_{i=1}^{N^t}\sum_{j=1}^{N^c}w_{ij}[(x^t_{ip})^k - (x^c_{jp})^k]\biggl| \leq \sigma_p^k\\
&\mbox{in Formulation 3: } &&\biggl|\sum_{i=1}^{N^t}\sum_{j=1}^{N^c}[(x^t_{ip})^k - (x^c_{jp})^k] \biggl|\leq \sigma_p^k\cdot z.
\end{align}
\end{subequations}
Balance can also be ensured by requiring quantiles of the distributions of the covariates in the two samples to not be too far apart. For the SATT, we can make this process efficient by employing the suggestion of \cite{Zubizarreta2012} and define, for a covariate $p$: $H_p = \{h_{p1}, \dots, h_{pk}\}$, a vector of proportions for each quantile of the distribution of $p$, and by computing $G_p^t = \{g_{p1}, \dots, g_{pk}\}$, the covariate quantile values in the treatment group corresponding to the proportion in set $H_p$. Then, for all covariates $p$ and quantiles, $k$ we can require:
\begin{subequations}
\begin{align}
\mbox{in Formulation 1: } & \biggl| h_{pk} - \frac{1}{N^t}\sum_{i=1}^{N^t}\sum_{j=1}^{N^c}w_{ij}\mathbb{I}(x_{ip}^tc\leq g_{pk}) \biggl|\leq \epsilon_{p}\\
\end{align}
For the FSATT, we impose quantile balance constraints by introducing a vector of $k$ desired quantile values for covariate $p$: $Q_p = \{q_{p1}, \dots, q_{pk}\}$ and create auxiliary variable $\mathbb{I}(x_{ip} \leq q_{pk})$, which takes value 1 if $x_{ip} \leq q_{pk}$ and 0 otherwise for all treated and untreated observations. Then, for all quantiles $k$ of each covariate $p$ we require:
\begin{align}
\mbox{in Formulation 2: } &\biggl|\frac{1}{M}\sum_{i=1}^{N^t}\sum_{j=1}^{N^c}w_{ij}\mathbb{I}(x_{ip}^t \leq q_{pk}) - \frac{1}{M}\sum_{i=1}^{N^t}\sum_{j=1}^{N^c}w_{ij}\mathbb{I}(x_{ip}^t\leq q_{pk})\biggl| \leq \epsilon_{p}\\
\mbox{in Formulation 3: } &\biggl|\sum_{i=1}^{N^t}\sum_{j=1}^{N^c}\mathbb{I}(x_{ip}^t \leq q_{pk}) - \sum_{i=1}^{N^t}\sum_{j=1}^{N^c}\mathbb{I}(x_{ip}^c\leq q_{pk}) \biggl|\leq \epsilon_{p}\cdot z.
\end{align}
\end{subequations}
Finally, we might want to constrain matches to be made exactly on a covariate, $p$. In order to do so we create the additional variable $\mathbb{I}(x_{ip} = x_{jp})$ and then require, for all treated units $i$ and control units $j$: $w_{ij} \leq \mathbb{I}(x_{ip}^t = x_{ip}^c)$.
Here, the decision variable $w_{ij}$ can be replaced by the transformed variable to estimate the FSATT with a variable number of units or by $u_{ij}$ to estimate the SATT with multiple matches. All the constraints that have absolute values in them must be linearized before being included in any integer programming formulation. To do so, it is sufficient to separate each constraint into two; i.e., $|aw| \leq b$ can be replaced by $aw \leq b$ and $-aw \leq b$.

\section{Applications: Additional Information}

To compute Matching Bounds for the replication exercise, we started from the matches obtained with either genetic matching $\W^{\GM}$ (Application 2) or pscore nearest neighbor-matching $\W^{\PS}$ (Appl. 1) and derived values for the balance constraints for the first three moments of the covariate distributions of the control group (treatment group for Appl. 1). These were computed using the formula in Eq. \eqref{Eq:Bal:Mom:F4}. Additionally, we also impose a caliper on the maximum distance that a control unit (treated in Appl. 1) can be from its respective matched treatment (control in Appl. 1). This caliper is equal to the distance between the treated (control) unit and its matched counterpart in the data matched by $\GM$ or $\PS$. All of these constraint values are scaled by a constant $\epsilon$ that we vary in order to simulate different levels of tolerance for our Matching Bounds. When $\epsilon$ is large, then we are more tolerant of matches that are far away from the original ones in terms of the balance constraints just defined. To compute Matching Bounds we solved the following mixed-integer program, for $A \in \{GM, PS\}$. and for $M = N^c$ in Appl. 1 and $M = N^t$ in Appl. 2:
\begin{align*}
\mbox{(Appl. 1) min/max } &\hat{\tau}^c = \frac{1}{M}\sum_{j=1}^{N^c}\sum_{i=1}^{N^t}y_i^cw_{ij}  - \bar{y}^c\\
\mbox{(Appl. 2) min/max } &\hat{\tau}^t = \bar{y}^t - \frac{1}{M}\sum_{i=1}^{N^t} \sum_{j=1}^{N^c}y_i^cw_{ij}
\end{align*}
\begin{align*}
\mbox{subject to: }&w_{ij} \in \{0,1\}, &&\quad i = 1,\dots,N^t, j=1,\dots,N^c \label{c1.1} \\
&\sum_{i=1}^{N^t}\sum_{j=1}^{N^c} w_{ij} = M, \\
&\biggl|\frac{1}{M}\sum_{i=1}^{N^t}\sum_{j=1}^{N^c}[x_{ip}^t - x_{jp}^c]w_{ij}\biggr| \leq \epsilon\sigma_{p1}^{A}, &&\quad p=1,\dots,P\\
&\biggl|\frac{1}{M}\sum_{i=1}^{N^t}\sum_{j=1}^{N^c}[(x_{ip}^t)^2 - (x_{jp}^c)^2]w_{ij}\biggr| \leq \epsilon\sigma_{p2}^{A},&&\quad p=1,\dots,P\\
&\biggl|\frac{1}{M}\sum_{i=1}^{N^t}\sum_{j=1}^{N^c}[(x_{ip}^t)^3 - (x_{jp}^c)^3]w_{ij}\biggr| \leq \epsilon\sigma_{p3}^{A},&&\quad p=1,\dots,P\\
&\sum_{i=1}^{N^t}\sum_{j=1}^{N^c} w_{ij}d_{ij}^A \leq \epsilon\sigma_{4ij}^A.
\end{align*}
We obtained $\W^+_{A}$ by solving the maximization problem and  $\W^-_A$ by solving the minimization problem. Once the Matching Bounds were obtained, we computed the corresponding TE estimates using the SATT estimator defined in~\eqref{Eq:ATTEst} (Appl. 2) or its correspondent SATC estimator (Appl. 2).

\subsection{KS tests of univariate covariate imbalance}
Here we report KS statistics of difference in distribution between the treated (control) groups and their matched counterparts in our bounds. Test statistics are reported in the top of each row and their respective p-value is reported below them in parentheses. P-values that are below 0.05 are bolded. Generally, in only a few extreme instances (i.e., when bounds are within twice the balance produced by the initial matches) we see a small portion of covariates with distributions statistically different from those of their matched counterparts. Generally, these tables suggest that all of our bounds achieve good balance with their matches. 

\begin{sidewaystable}[ht]
\centering
\resizebox{\textwidth}{!}{%
\begin{tabular}{rrllllllllllllll}
  \hline
 Epsilon & Bound & pscore & industry & income & crops & land & home & indstry & community& motorable & maoist & children & education & gender & age \\ 
 & & & & seized & seized & destroyed & destroyed & & & roads & & & & & \\
  \hline
  0 & unmatched & 0.285 & 0.037 & 0.134 & 0.073 & 0.016 & 0.093 & 0.028 & 0.212 & 0.068 & 0.053 & 0.045 & 0.089 & 0.032 & 0 \\ 
   &  & (0) & (0.996) & (0.026) & (0.541) & (1) & (0.245) & (1) & (0) & (0.628) & (0.886) & (0.968) & (0.289) & (1) & (NA) \\ 
  0 & original & 0.453 & 0 & 0 & 0 & 0 & 0 & 0 & 0 & 0 & 0 & 0 & 0 & 0 & 0 \\ 
   &  & (0) & (1) & (1) & (1) & (1) & (1) & (1) & (1) & (1) & (1) & (1) & (1) & (1) & (NA) \\ 
  1 & max & 0.015 & 0.023 & 0.053 & 0.053 & 0.038 & 0 & 0.023 & 0 & 0.068 & 0.053 & 0.09 & 0.09 & 0 & 0.075 \\ 
   &  & (1) & (1) & (0.993) & (0.993) & (1) & (1) & (1) & (1) & (0.921) & (0.993) & (0.651) & (0.651) & (1) & (0.846) \\ 
  1 & min & 0.015 & 0.023 & 0.053 & 0.053 & 0.038 & 0 & 0.023 & 0 & 0.068 & 0.053 & 0.09 & 0.09 & 0 & 0.075 \\ 
   &  & (1) & (1) & (0.993) & (0.993) & (1) & (1) & (1) & (1) & (0.921) & (0.993) & (0.651) & (0.651) & (1) & (0.846) \\ 
  1.1 & max & 0.015 & 0.023 & 0.053 & 0.053 & 0.038 & 0 & 0.023 & 0 & 0.068 & 0.053 & 0.09 & 0.09 & 0 & 0.075 \\ 
    &  & (1) & (1) & (0.993) & (0.993) & (1) & (1) & (1) & (1) & (0.921) & (0.993) & (0.651) & (0.651) & (1) & (0.846) \\ 
   1.1 & min & 0.015 & 0.023 & 0.053 & 0.053 & 0.038 & 0 & 0.023 & 0 & 0.068 & 0.053 & 0.09 & 0.09 & 0 & 0.075 \\ 
    &  & (1) & (1) & (0.993) & (0.993) & (1) & (1) & (1) & (1) & (0.921) & (0.993) & (0.651) & (0.651) & (1) & (0.846) \\ 
   1.2 & max & 0.015 & 0.023 & 0.038 & 0.023 & 0.03 & 0 & 0.023 & 0 & 0.068 & 0.053 & 0.113 & 0.098 & 0 & 0.053 \\ 
    &  & (1) & (1) & (1) & (1) & (1) & (1) & (1) & (1) & (0.921) & (0.993) & (0.366) & (0.549) & (1) & (0.993) \\ 
   1.2 & min & 0.015 & 0.023 & 0.038 & 0.045 & 0.045 & 0 & 0.023 & 0 & 0.06 & 0.045 & 0.113 & 0.083 & 0 & 0.075 \\ 
    &  & (1) & (1) & (1) & (0.999) & (0.999) & (1) & (1) & (1) & (0.97) & (0.999) & (0.366) & (0.753) & (1) & (0.846) \\ 
   1.3 & max & 0.015 & 0.023 & 0.038 & 0.03 & 0.023 & 0 & 0.023 & 0 & 0.053 & 0.068 & 0.113 & 0.105 & 0 & 0.053 \\ 
    &  & (1) & (1) & (1) & (1) & (1) & (1) & (1) & (1) & (0.993) & (0.921) & (0.366) & (0.453) & (1) & (0.993) \\ 
   1.3 & min & 0.015 & 0.023 & 0.03 & 0.038 & 0.038 & 0 & 0.023 & 0 & 0.053 & 0.053 & 0.105 & 0.083 & 0 & 0.068 \\ 
    &  & (1) & (1) & (1) & (1) & (1) & (1) & (1) & (1) & (0.993) & (0.993) & (0.453) & (0.753) & (1) & (0.921) \\ 
   1.4 & max & 0.015 & 0.03 & 0.038 & 0.015 & 0.015 & 0 & 0.023 & 0 & 0.068 & 0.06 & 0.128 & 0.083 & 0 & 0.045 \\ 
    &  & (1) & (1) & (1) & (1) & (1) & (1) & (1) & (1) & (0.921) & (0.97) & (0.227) & (0.753) & (1) & (0.999) \\ 
   1.4 & min & 0.015 & 0.023 & 0.023 & 0.03 & 0.053 & 0 & 0.015 & 0 & 0.06 & 0.045 & 0.098 & 0.075 & 0 & 0.068 \\ 
    &  & (1) & (1) & (1) & (1) & (0.993) & (1) & (1) & (1) & (0.97) & (0.999) & (0.549) & (0.846) & (1) & (0.921) \\ 
   1.5 & max & 0.015 & 0.03 & 0.038 & 0.023 & 0.038 & 0 & 0.023 & 0 & 0.06 & 0.06 & 0.128 & 0.105 & 0 & 0.045 \\ 
    &  & (1) & (1) & (1) & (1) & (1) & (1) & (1) & (1) & (0.97) & (0.97) & (0.227) & (0.453) & (1) & (0.999) \\ 
   1.5 & min & 0.015 & 0.03 & 0.03 & 0.023 & 0.053 & 0 & 0.023 & 0 & 0.068 & 0.03 & 0.105 & 0.075 & 0 & 0.083 \\ 
    &  & (1) & (1) & (1) & (1) & (0.993) & (1) & (1) & (1) & (0.921) & (1) & (0.453) & (0.846) & (1) & (0.753) \\ 
   1.6 & max & 0.015 & 0.03 & 0.045 & 0.015 & 0.015 & 0 & 0.023 & 0 & 0.09 & 0.053 & 0.12 & 0.09 & 0 & 0.06 \\ 
    &  & (1) & (1) & (0.999) & (1) & (1) & (1) & (1) & (1) & (0.651) & (0.993) & (0.291) & (0.651) & (1) & (0.97) \\ 
   1.6 & min & 0.023 & 0.03 & 0.053 & 0.023 & 0.03 & 0 & 0.023 & 0 & 0.083 & 0.045 & 0.083 & 0.015 & 0.008 & 0.053 \\ 
    &  & (1) & (1) & (0.993) & (1) & (1) & (1) & (1) & (1) & (0.753) & (0.999) & (0.753) & (1) & (1) & (0.993) \\ 
   1.7 & max & 0.015 & 0.03 & 0.038 & 0.023 & 0 & 0 & 0.023 & 0 & 0.083 & 0.083 & 0.113 & 0.098 & 0 & 0.045 \\ 
    &  & (1) & (1) & (1) & (1) & (1) & (1) & (1) & (1) & (0.753) & (0.753) & (0.366) & (0.549) & (1) & (0.999) \\ 
   1.7 & min & 0.015 & 0.03 & 0.045 & 0.03 & 0.03 & 0 & 0.023 & 0 & 0.075 & 0.038 & 0.098 & 0.068 & 0 & 0.068 \\ 
    &  & (1) & (1) & (0.999) & (1) & (1) & (1) & (1) & (1) & (0.846) & (1) & (0.549) & (0.921) & (1) & (0.921) \\ 
   1.8 & max & 0.015 & 0.023 & 0.053 & 0.008 & 0.023 & 0 & 0.015 & 0 & 0.038 & 0.075 & 0.09 & 0.105 & 0 & 0.068 \\ 
    &  & (1) & (1) & (0.993) & (1) & (1) & (1) & (1) & (1) & (1) & (0.846) & (0.651) & (0.453) & (1) & (0.921) \\ 
   1.8 & min & 0.015 & 0.03 & 0.053 & 0.023 & 0.045 & 0 & 0.03 & 0 & 0.075 & 0.015 & 0.075 & 0.053 & 0 & 0.06 \\ 
    &  & (1) & (1) & (0.993) & (1) & (0.999) & (1) & (1) & (1) & (0.846) & (1) & (0.846) & (0.993) & (1) & (0.97) \\ 
   1.9 & max & 0.015 & 0.023 & 0.06 & 0.015 & 0.015 & 0 & 0.023 & 0 & 0.06 & 0.075 & 0.105 & 0.083 & 0 & 0.053 \\ 
    &  & (1) & (1) & (0.97) & (1) & (1) & (1) & (1) & (1) & (0.97) & (0.846) & (0.453) & (0.753) & (1) & (0.993) \\ 
   1.9 & min & 0.015 & 0.023 & 0.068 & 0.03 & 0.045 & 0 & 0.03 & 0 & 0.06 & 0.008 & 0.09 & 0.038 & 0 & 0.045 \\ 
    &  & (1) & (1) & (0.921) & (1) & (0.999) & (1) & (1) & (1) & (0.97) & (1) & (0.651) & (1) & (1) & (0.999) \\ 
   2 & max & 0.015 & 0.023 & 0.053 & 0.008 & 0.008 & 0 & 0.015 & 0 & 0.068 & 0.075 & 0.098 & 0.083 & 0 & 0.06 \\ 
    &  & (1) & (1) & (0.993) & (1) & (1) & (1) & (1) & (1) & (0.921) & (0.846) & (0.549) & (0.753) & (1) & (0.97) \\ 
   2 & min & 0.023 & 0.03 & 0.06 & 0.008 & 0.045 & 0 & 0.023 & 0 & 0.075 & 0 & 0.098 & 0.06 & 0 & 0.06 \\ 
    &  & (1) & (1) & (0.97) & (1) & (0.999) & (1) & (1) & (1) & (0.846) & (1) & (0.549) & (0.97) & (1) & (0.97) \\ 
   \hline
\end{tabular}}
\end{sidewaystable}

\begin{sidewaystable}[ht]
\centering
\resizebox{\textwidth}{!}{%
\begin{tabular}{rrlllllllllllllllll}
  \hline
 Epsilon & Bound & pscore & resvol & totvol & bopgdp & resgdp & gdpgrow & useimfcr & gnpcap & surveil & regnorm & flexible & tradedep & univers & military & termlim & parli & lastrest \\ 
  \hline
  0 & unmatched & 0.29 & 0.193 & 0.1 & 0.103 & 0.203 & 0.107 & 0.122 & 0.138 & 0.089 & 0.16 & 0.048 & 0.18 & 0.097 & 0.112 & 0.07 & 0.134 & 0.11 \\ 
   &  & (0) & (0) & (0.015) & (0.012) & (0) & (0.007) & (0.001) & (0) & (0.04) & (0) & (0.63) & (0) & (0.021) & (0.004) & (0.182) & (0) & (0.005) \\ 
  0 & original & 0.061 & 0.048 & 0.061 & 0.106 & 0.051 & 0.044 & 0.048 & 0.065 & 0 & 0.041 & 0.007 & 0.061 & 0.034 & 0.041 & 0.031 & 0.058 & 0.038 \\ 
   &  & (0.638) & (0.892) & (0.638) & (0.075) & (0.837) & (0.935) & (0.892) & (0.569) & (1) & (0.967) & (1) & (0.638) & (0.996) & (0.967) & (0.999) & (0.707) & (0.986) \\ 
  1 & max & 0.061 & 0.048 & 0.061 & 0.106 & 0.051 & 0.044 & 0.048 & 0.065 & 0 & 0.041 & 0.007 & 0.061 & 0.034 & 0.041 & 0.031 & 0.058 & 0.038 \\ 
   &  & (0.638) & (0.892) & (0.638) & (0.075) & (0.837) & (0.935) & (0.892) & (0.569) & (1) & (0.967) & (1) & (0.638) & (0.996) & (0.967) & (0.999) & (0.707) & (0.986) \\ 
  1 & min & 0.061 & 0.048 & 0.061 & 0.106 & 0.051 & 0.044 & 0.048 & 0.065 & 0 & 0.041 & 0.007 & 0.061 & 0.034 & 0.041 & 0.031 & 0.058 & 0.038 \\ 
   &  & (0.638) & (0.892) & (0.638) & (0.075) & (0.837) & (0.935) & (0.892) & (0.569) & (1) & (0.967) & (1) & (0.638) & (0.996) & (0.967) & (0.999) & (0.707) & (0.986) \\ 
  1.1 & max & 0.055 & 0.038 & 0.065 & 0.096 & 0.065 & 0.044 & 0.048 & 0.061 & 0 & 0.041 & 0.003 & 0.058 & 0.031 & 0.034 & 0.031 & 0.051 & 0.031 \\ 
   &  & (0.775) & (0.986) & (0.569) & (0.138) & (0.569) & (0.935) & (0.892) & (0.638) & (1) & (0.967) & (1) & (0.707) & (0.999) & (0.996) & (0.999) & (0.837) & (0.999) \\ 
  1.1 & min & 0.061 & 0.048 & 0.055 & 0.082 & 0.044 & 0.041 & 0.044 & 0.055 & 0 & 0.061 & 0.003 & 0.051 & 0.031 & 0.041 & 0.034 & 0.051 & 0.048 \\ 
   &  & (0.638) & (0.892) & (0.775) & (0.279) & (0.935) & (0.967) & (0.935) & (0.775) & (1) & (0.638) & (1) & (0.837) & (0.999) & (0.967) & (0.996) & (0.837) & (0.892) \\ 
  1.2 & max & 0.055 & 0.041 & 0.065 & 0.082 & 0.055 & 0.044 & 0.041 & 0.058 & 0 & 0.041 & 0.003 & 0.061 & 0.048 & 0.031 & 0.044 & 0.031 & 0.041 \\ 
   &  & (0.775) & (0.967) & (0.569) & (0.279) & (0.775) & (0.935) & (0.967) & (0.707) & (1) & (0.967) & (1) & (0.638) & (0.892) & (0.999) & (0.935) & (0.999) & (0.967) \\ 
  1.2 & min & 0.065 & 0.058 & 0.061 & 0.102 & 0.044 & 0.044 & 0.041 & 0.061 & 0 & 0.055 & 0.003 & 0.051 & 0.027 & 0.048 & 0.038 & 0.051 & 0.055 \\ 
   &  & (0.569) & (0.707) & (0.638) & (0.093) & (0.935) & (0.935) & (0.967) & (0.638) & (1) & (0.775) & (1) & (0.837) & (1) & (0.892) & (0.986) & (0.837) & (0.775) \\ 
  1.3 & max & 0.065 & 0.044 & 0.065 & 0.096 & 0.072 & 0.038 & 0.027 & 0.041 & 0 & 0.041 & 0.007 & 0.072 & 0.034 & 0.027 & 0.031 & 0.051 & 0.068 \\ 
   &  & (0.569) & (0.935) & (0.569) & (0.138) & (0.439) & (0.986) & (1) & (0.967) & (1) & (0.967) & (1) & (0.439) & (0.996) & (1) & (0.999) & (0.837) & (0.502) \\ 
  1.3 & min & 0.055 & 0.092 & 0.068 & 0.119 & 0.058 & 0.041 & 0.055 & 0.048 & 0 & 0.061 & 0.007 & 0.092 & 0.044 & 0.041 & 0.031 & 0.061 & 0.058 \\ 
   &  & (0.775) & (0.166) & (0.502) & \textbf{(0.031)} & (0.707) & (0.967) & (0.775) & (0.892) & (1) & (0.638) & (1) & (0.166) & (0.935) & (0.967) & (0.999) & (0.638) & (0.707) \\ 
  1.4 & max & 0.075 & 0.068 & 0.068 & 0.126 & 0.075 & 0.051 & 0.027 & 0.051 & 0 & 0.048 & 0.007 & 0.082 & 0.055 & 0.027 & 0.038 & 0.044 & 0.085 \\ 
   &  & (0.381) & (0.502) & (0.502) & \textbf{(0.019)} & (0.381) & (0.837) & (1) & (0.837) & (1) & (0.892) & (1) & (0.279) & (0.775) & (1) & (0.986) & (0.935) & (0.237) \\ 
  1.4 & min & 0.055 & 0.082 & 0.078 & 0.119 & 0.068 & 0.051 & 0.048 & 0.055 & 0 & 0.048 & 0.007 & 0.072 & 0.051 & 0.038 & 0.031 & 0.058 & 0.096 \\ 
   &  & (0.775) & (0.279) & (0.327) & \textbf{(0.031)} & (0.502) & (0.837) & (0.892) & (0.775) & (1) & (0.892) & (1) & (0.439) & (0.837) & (0.986) & (0.999) & (0.707) & (0.138) \\ 
  1.5 & max & 0.065 & 0.041 & 0.072 & 0.078 & 0.051 & 0.055 & 0.027 & 0.065 & 0 & 0.038 & 0.003 & 0.072 & 0.031 & 0.055 & 0.055 & 0.061 & 0.082 \\ 
   &  & (0.569) & (0.967) & (0.439) & (0.327) & (0.837) & (0.775) & (1) & (0.569) & (1) & (0.986) & (1) & (0.439) & (0.999) & (0.775) & (0.775) & (0.638) & (0.279) \\ 
  1.5 & min & 0.051 & 0.092 & 0.089 & 0.123 & 0.048 & 0.041 & 0.055 & 0.058 & 0 & 0.075 & 0.003 & 0.065 & 0.048 & 0.051 & 0.031 & 0.051 & 0.133 \\ 
   &  & (0.837) & (0.166) & (0.199) & \textbf{(0.024)} & (0.892) & (0.967) & (0.775) & (0.707) & (1) & (0.381) & (1) & (0.569) & (0.892) & (0.837) & (0.999) & (0.837) & \textbf{(0.011)} \\ 
  1.6 & max & 0.061 & 0.058 & 0.078 & 0.109 & 0.078 & 0.072 & 0.01 & 0.051 & 0 & 0.038 & 0.01 & 0.065 & 0.027 & 0.048 & 0.034 & 0.038 & 0.102 \\ 
   &  & (0.638) & (0.707) & (0.327) & (0.061) & (0.327) & (0.439) & (1) & (0.837) & (1) & (0.986) & (1) & (0.569) & (1) & (0.892) & (0.996) & (0.986) & (0.093) \\ 
  1.6 & min & 0.055 & 0.099 & 0.102 & 0.106 & 0.061 & 0.044 & 0.068 & 0.061 & 0 & 0.055 & 0 & 0.065 & 0.041 & 0.058 & 0.038 & 0.061 & 0.113 \\ 
   &  & (0.775) & (0.113) & (0.093) & (0.075) & (0.638) & (0.935) & (0.502) & (0.638) & (1) & (0.775) & (1) & (0.569) & (0.967) & (0.707) & (0.986) & (0.638) & \textbf{(0.049)} \\ 
  1.7 & max & 0.058 & 0.055 & 0.068 & 0.065 & 0.058 & 0.061 & 0.02 & 0.068 & 0 & 0.051 & 0.003 & 0.065 & 0.031 & 0.051 & 0.055 & 0.061 & 0.137 \\ 
   &  & (0.707) & (0.775) & (0.502) & (0.569) & (0.707) & (0.638) & (1) & (0.502) & (1) & (0.837) & (1) & (0.569) & (0.999) & (0.837) & (0.775) & (0.638) & \textbf{(0.009)} \\ 
  1.7 & min & 0.065 & 0.085 & 0.089 & 0.119 & 0.068 & 0.051 & 0.068 & 0.061 & 0 & 0.061 & 0.007 & 0.085 & 0.041 & 0.044 & 0.041 & 0.072 & 0.137 \\ 
   &  & (0.569) & (0.237) & (0.199) & \textbf{(0.031)} & (0.502) & (0.837) & (0.502) & (0.638) & (1) & (0.638) & (1) & (0.237) & (0.967) & (0.935) & (0.967) & (0.439) & \textbf{(0.009)} \\ 
  1.8 & max & 0.078 & 0.085 & 0.078 & 0.116 & 0.058 & 0.072 & 0.027 & 0.072 & 0 & 0.065 & 0.007 & 0.065 & 0.041 & 0.072 & 0.038 & 0.031 & 0.13 \\ 
   &  & (0.327) & (0.237) & (0.327) & \textbf{(0.039)} & (0.707) & (0.439) & (1) & (0.439) & (1) & (0.569) & (1) & (0.569) & (0.967) & (0.439) & (0.986) & (0.999) & \textbf{(0.014)} \\ 
  1.8 & min & 0.065 & 0.102 & 0.102 & 0.109 & 0.092 & 0.058 & 0.061 & 0.038 & 0 & 0.061 & 0.01 & 0.078 & 0.034 & 0.065 & 0.061 & 0.055 & 0.133 \\ 
   &  & (0.569) & (0.093) & (0.093) & (0.061) & (0.166) & (0.707) & (0.638) & (0.986) & (1) & (0.638) & (1) & (0.327) & (0.996) & (0.569) & (0.638) & (0.775) & \textbf{(0.011)} \\ 
  1.9 & max & 0.055 & 0.055 & 0.068 & 0.075 & 0.102 & 0.061 & 0.007 & 0.072 & 0 & 0.082 & 0.01 & 0.065 & 0.065 & 0.075 & 0.061 & 0.048 & 0.123 \\ 
   &  & (0.775) & (0.775) & (0.502) & (0.381) & (0.093) & (0.638) & (1) & (0.439) & (1) & (0.279) & (1) & (0.569) & (0.569) & (0.381) & (0.638) & (0.892) & \textbf{(0.024)} \\ 
  1.9 & min & 0.061 & 0.078 & 0.096 & 0.116 & 0.065 & 0.082 & 0.068 & 0.041 & 0 & 0.085 & 0.003 & 0.089 & 0.048 & 0.065 & 0.055 & 0.082 & 0.157 \\ 
   &  & (0.638) & (0.327) & (0.138) & \textbf{(0.039)} & (0.569) & (0.279) & (0.502) & (0.967) & (1) & (0.237) & (1) & (0.199) & (0.892) & (0.569) & (0.775) & (0.279) & \textbf{(0.001)} \\ 
  2 & max & 0.061 & 0.082 & 0.058 & 0.075 & 0.061 & 0.065 & 0.007 & 0.106 & 0 & 0.085 & 0.014 & 0.061 & 0.048 & 0.051 & 0.055 & 0.034 & 0.167 \\ 
   &  & (0.638) & (0.279) & (0.707) & (0.381) & (0.638) & (0.569) & (1) & \textbf{(0.075)} & (1) & (0.237) & (1) & (0.638) & (0.892) & (0.837) & (0.775) & (0.996) & \textbf{(0.001)} \\ 
  2 & min & 0.072 & 0.096 & 0.099 & 0.096 & 0.072 & 0.044 & 0.075 & 0.051 & 0 & 0.075 & 0.003 & 0.068 & 0.041 & 0.061 & 0.048 & 0.061 & 0.164 \\ 
   &  & (0.439) & (0.138) & (0.113) & (0.138) & (0.439) & (0.935) & (0.381) & (0.837) & (1) & (0.381) & (1) & (0.502) & (0.967) & (0.638) & (0.892) & (0.638) & \textbf{(0.001)} \\ 
   \hline
\end{tabular}}
\end{sidewaystable}

\end{document}